\documentclass{article}

\usepackage[final]{neurips_2025}

\usepackage[utf8]{inputenc} 
\usepackage[T1]{fontenc}    
\usepackage{hyperref}       
\usepackage{url}            
\usepackage{booktabs}       
\usepackage{amsfonts}       
\usepackage{nicefrac}       
\usepackage{microtype}      
\usepackage{xcolor}         
\usepackage{booktabs}
\usepackage{enumitem}
\usepackage{graphicx}
\usepackage{subfigure}
\usepackage{amsmath, amssymb, amsthm, bbm, mathrsfs, mathtools}
\usepackage{algorithmic}
\usepackage{algorithm}
\usepackage{multirow}
\mathtoolsset{showonlyrefs=true}

\usepackage{etoolbox}
\usepackage{tabularx, booktabs, graphicx}

\makeatletter
\newcommand{\ostar}{\mathbin{\mathpalette\make@circled\star}}
\newcommand{\make@circled}[2]{%
  \ooalign{$\m@th#1\smallbigcirc{#1}$\cr\hidewidth$\m@th#1#2$\hidewidth\cr}%
}
\newcommand{\smallbigcirc}[1]{%
  \vcenter{\hbox{\scalebox{0.77778}{$\m@th#1\bigcirc$}}}%
}
\makeatother

\newcommand{\lt}{\left}
\newcommand{\rt}{\right}

\newcommand{\commenting}[1]{}

\renewcommand{\hat}{\widehat}

\newcommand{\E}[1]{{\bbE\left\{#1\right\}}}
\newcommand{\Prob}[1]{{\bbP\left\{#1\right\}}}

\newcommand{\innerprod}[2]{\left\langle#1\, ,\, #2\right\rangle}

\ifdef{\see}{\renewcommand{\see}[1]{\text{ (#1)}}}{\newcommand{\see}[1]{\text{ (#1)}}}

\newcommand{\diag}[1]{\mathrm{Diag}\left\{#1\right\}}

\def\boxit#1{\vbox{\hrule\hbox{\vrule\kern6pt\vbox{\kern6pt#1\kern6pt}\kern6pt\vrule}\hrule}}

\newcolumntype{P}[1]{>{\centering\arraybackslash}p{#1}}
\newcolumntype{M}[1]{>{\centering\arraybackslash}m{#1}}
\newcolumntype{L}[1]{>{\raggedright\arraybackslash}m{#1}}

\DeclareMathOperator*{\argmin}{arg\,min}

\newcommand{\cB}{{\mathcal{B}}}

\newcommand{\cX}{{\mathcal{X}}}

\newcommand{\cZ}{{\mathcal{Z}}}
\newcommand{\cC}{{\mathcal{C}}}

\newcommand{\cN}{{\mathcal{N}}}

\newcommand{\cK}{{\mathcal{K}}}

\newcommand{\fS}{{\mathfrak{S}}}

\newcommand{\bbH}{{\mathbb{H}}}

\newcommand{\bbP}{{\mathbb{P}}}

\newcommand{\bbR}{{\mathbb{R}}}

\newcommand{\bbB}{{\mathbb{B}}}
\newcommand{\bbE}{{\mathbb{E}}}

\newcommand{\bone}{{\boldsymbol{1}}}
\newcommand{\bzero}{{\boldsymbol{0}}}

\newcommand{\bg}{{\boldsymbol{g}}}
\newcommand{\by}{{\boldsymbol{y}}}

\newcommand{\bh}{{\boldsymbol{h}}}

\newcommand{\bx}{{\boldsymbol{x}}}
\newcommand{\bz}{{\boldsymbol{z}}}

\newcommand{\bR}{{\boldsymbol{R}}}
\newcommand{\bS}{{\boldsymbol{S}}}

\newcommand{\bU}{{\boldsymbol{U}}}
\newcommand{\bV}{{\boldsymbol{V}}}

\newcommand{\bX}{{\boldsymbol{X}}}

\newcommand{\bZ}{{\boldsymbol{Z}}}

\newcommand{\bepsilon}{{\boldsymbol{\epsilon}}}

\newcommand{\bTheta}{{\boldsymbol{\Theta}}}
\newcommand{\bGamma}{{\boldsymbol{\Gamma}}}

\newcommand{\bDelta}{{\boldsymbol{\Delta}}}
\newcommand{\bOmega}{{\boldsymbol{\Omega}}}
\newcommand{\bLambda}{{\boldsymbol{\Lambda}}}

\newcommand{\halpha}{{\hat{\alpha}}}

\newcommand{\hbeta}{{\hat{\beta}}}

\newcommand{\hepsilon}{{\hat{\epsilon}}}

\newcommand{\hatm}{{\hat{m}}}

\theoremstyle{plain}
\newtheorem{theorem}{Theorem}[section]
\newtheorem{proposition}[theorem]{Proposition}

\theoremstyle{definition}
\newtheorem{definition}[theorem]{Definition}

\theoremstyle{remark}

\newcommand{\Kg}{\cK_g}
\newcommand{\Kh}{\cK_h}

\newcommand{\bUc}{\bU^{\textsc{c}}}
\newcommand{\bUr}{\bU^{\textsc{r}}}
\newcommand{\bVc}{\bV^{\textsc{c}}}
\newcommand{\bVr}{\bV^{\textsc{r}}}

\newcommand{\bRc}{\bR^{\textsc{c}}}
\newcommand{\bRr}{\bR^{\textsc{r}}}
\newcommand{\bSc}{\bS^{\textsc{c}}}
\newcommand{\bSr}{\bS^{\textsc{r}}}

\title{Leveraging semantic similarity for experimentation with AI-generated treatments}

\author{%
  Lei Shi \\
  University of California, Berkeley\\
  Berkeley, CA 94720 \\
  \texttt{leishi@berkeley.edu} \\
  \And
  David Arbour \\
  Adobe Research \\
  San Jose, CA 95110\\
  \texttt{arbour@adobe.com} \\
  \And
  Raghavendra Addanki\\
  Adobe Research \\
  San Jose, CA 95110\\
  \texttt{raddanki@adobe.com} \\
  \And
  Ritwik Sinha \\
  Adobe Research \\
  San Jose, CA 95110\\
  \texttt{risinha@adobe.com} \\
  \And
  Avi Feller \\
  University of California, Berkeley\\
  Berkeley, CA 94720 \\
  \texttt{afeller@berkeley.edu}
}

\begin{document}

\maketitle

\begin{abstract}
Large Language Models (LLMs) enable a new form of digital experimentation where treatments combine human and model-generated content in increasingly sophisticated ways. The main methodological challenge in this setting is representing these high-dimensional treatments without losing their semantic meaning or rendering analysis intractable. Here we address this problem by focusing on learning low-dimensional representations that capture the underlying structure of such treatments. These representations enable downstream applications such as guiding generative models to produce meaningful treatment variants and facilitating adaptive assignment in online experiments. We propose double kernel representation learning, which models the causal effect through the inner product of kernel-based representations of treatments and user covariates. We develop an alternating-minimization algorithm that learns these representations efficiently from data and provide convergence guarantees under a low-rank factor model. As an application of this framework, we introduce an adaptive design strategy for online experimentation and demonstrate the method's effectiveness through numerical experiments.
\end{abstract}

\section{Introduction}
\label{sec:intro}
Randomized experiments are a cornerstone of measuring the efficacy of content (e.g., messaging, imagery) in modern industry         ~\citep[e.g.,][]{sun2021creating,grimmer2014green,leung2017hotel} and social science~\citep[e.g.,][]{gerber2018comparative,green2023using,salovey2004field} contexts. 
Historically, generating the content itself has been the core challenge.
The advent of Large Language Models (LLMs) and other generative models involving text, images, and videos \citep{golkab2023impact, baek2023digital, kumar2024generative} in recent years has dramatically decreased this design burden.
By integrating human creativity with model-generated content, generative models are transforming digital experimentation and streamlining treatment generation.

The ability to generate many content variants for experiments, however, has come with new statistical challenges, especially since the corresponding experiment populations have largely stayed fixed. 
As a result, practitioners often face the choice of either running underpowered experiments or leaving many treatment variations unexamined. In this paper, we explore an alternative: to improve statistical efficiency by 
explicitly incorporating semantic similarity from treatment embeddings (numeric representations of the treatments).
This raises two key technical challenges: (i) Effectively leveraging semantic information for better estimation and experimental designs, and
(ii) Developing a framework that appropriately captures the relationship between generated variants and user attributes. Dealing with these challenges can facilitate many downstream applications, such as guiding generative models to generate treatments and assisting adaptive treatment assignment in online experiments.

In this work, we propose a kernel-based representation learning approach that addresses these challenges. Our framework integrates both treatment embeddings and user covariate information to estimate the causal effects of LLM-generated variants. 
Specifically, our core contributions are: 

\begin{enumerate}
\item A double-kernel representation learning framework in which the treatment effect is the inner product of low-dimensional representations of treatment embeddings and user covariates.

\item A computationally efficient alternating minimization-style algorithm to learn these unknown representations. This naturally extends to an adaptive algorithm that assigns treatments to users in an online manner. 

\item Theoretical guarantees, including convergence rates for treatment effect estimation and a sublinear regret bounds of an algorithm for the adaptive experimentation setting.
\end{enumerate}

\section{Related Work and Background}\label{sec:related-work}
Our work builds on several key threads.

\textbf{Causal inference with structured treatments.}
There is an extensive literature on causal inference with text; see \cite{feder2022causal} for a comprehensive survey. Prior work has largely focused on the challenging problem of learning latent treatments from images or text~\citep{fong2023causal, egami2022make}. 
We sidestep several of these issues by taking the generated texts as fixed treatments and only focusing on using semantic information to improve efficiency. 
Many recent works have started exploring textual causal inference with LLMs, such as \citet{imai2024causal, tierney2025design}. However, they are not touching the problem of personalized treatment effect estimation and adaptive experimentation.

\textbf{High-dimensional and continuous treatments.} 
Our work builds most directly on recent results on heterogeneous treatment effects with high-dimensional and continuous treatments. \cite{kaddour2021causal} extended the R-learner approach of \cite{nie2021quasi} to a setting with unstructured treatments. 
\cite{liu2024encoding} studied continuous treatment effect estimation by modeling dependences among treatment and responses using covariate embeddings from generative models. 
\cite{singh2024kernel} proposed estimators based on kernel ridge regression for nonparametric causal functions such as dose, heterogeneous, and incremental response curves. 

More generally, there has been substantial recent work on experimentation with high-dimensional or continuous treatments, especially in the structured bandit problems, such as linear bandits~\citep{rusmevichientong2010linearly, hao2020high, oh2021sparsity, komiyama2024high}, generalized linear bandits \citep{filippi2010parametric, li2017provably}, low-rank bandits \citep{lu2021low, jun2019bilinear}, among others.


\textbf{Matrix completion for causal inference.}
There is a substantial literature on decomposing a function (e.g., treatment effect) based on matrix factorization, such as~\cite{abernethy2006low, fithian2013flexible, mao2019matrix, zhou2012kernelized, athey2021matrix}.  Most relevant for our work,~\cite{jin2022matrix} propose low-rank matrix completion when covariate information is available. Importantly,~\cite{jin2022matrix} only have covariate information available for the treatment-user pair, which differs from our setting in which we have both treatment embeddings and user attributes available.


\textbf{Kernel methods.} We build on the expansive causal inference literature that leveraging kernel methods: kernel-based methods that incorporate treatment embeddings for experimental design and analysis offer several benefits. First, kernel-based methods allow for flexible response surface modeling, $f\in\bbH_\cK$, going beyond simpler parametric formulations such as linear models \citep{rusmevichientong2010linearly, lu2010contextual}. Second, kernel-based methods provide adaptive design frameworks through tools including kernel bandits \citep{srinivas2009gaussian, valko2013finite} and Bayes optimization \citep{frazier2018tutorial, martinez2014bayesopt, ignatiadis2019covariate, dimmery2019shrinkage}. 
Third, kernel-based methods have well-established statistical guarantees, including prediction/estimation error analysis \citep{mendelson2010regularization, wainwright2019high} and regret analysis \citep{srinivas2009gaussian,  frazier2018tutorial, chowdhury2017kernelized, valko2013finite, vakili2023delayed}. 

\section{Problem description}\label{sec:problem}

\subsection{Setup}

\paragraph{Notation.} 
We first introduce notation. 
Throughout, for an integer $k$, $[k]$ denotes the set $\{1,\dots,k\}$. For a matrix $\bGamma$, $\|\bGamma\|_F$ is the Frobenius norm, $\|\bGamma\|_*$ is the nuclear norm, and $\|\bGamma\|_\infty$ is the element-wise infinity norm. $\rho_{\min}(\bGamma)$ and $\rho_{\max}(\bGamma)$ denote the smallest and largest singular values, respectively. 

\paragraph{Causal inference problem.}
Evaluating many content variants generated by LLMs creates challenges for classical A/B testing. We formalize these challenges under a causal inference framework. For a user with covariate $x \in \cX$, we posit the following structural model for the potential outcome under the treatment $z\in\cZ$:
\begin{align}\label{eqn:fzx}
    y(z) = f(z, x) + \epsilon, \quad z\in \cZ, \quad x\in\cX. 
\end{align}
Here $f$ is an unknown smooth function, and $\epsilon$ is mean-zero random noise. $\cZ$ and $\cX$ can be either finite or continuous spaces, representing the treatment and user covariate space, respectively. The structural model \eqref{eqn:fzx} implicitly encodes SUTVA: (i) no interference between units, meaning one units' treatments or outcomes do not affect another unit's potential outcomes; and (ii) no hidden version of treatments, meaning the realized outcome is the potential outcome at the assigned treatment.

For the estimation problem, we are interested in comparing a new treatment $z$ with a control variant $z_0$, for a user with covariates $x$. The conditional average treatment effect is then defined as:
\begin{align}
    \tau(z,x) = f(z,x) - f(z_0,x).
\end{align}
When $z$ is binary, i.e., $z\in\{z_0,z_1\}$, $\tau(z,x)$ is the standard conditional average treatment effect.

Learning $\tau(z,x)$ efficiently from data is crucial for both estimation and experimentation. For traditional A/B testing, the treatment space $\cZ$ involves a finite number of variants, which can be fully explored by assigning each element in $\cZ$ to a sufficient number of users. However, treating variants as separate discrete elements in LLM-powered experiments leads to two issues: (i) experiments are hugely underpowered since it is difficult to fully explore the treatment space $\cZ$ due to the large number of variants; (ii) the experiments ignore the semantic information in the variants, which encodes similarity between arms and could deliver important information about the outcome function. The two questions motivate the idea of leveraging treatment embeddings to explore the similarity between treatment arms and pool information across variants to improve statistical efficiency. 
We will now describe each of these problems in turn.

\subsection{Warmup: learning with treatment embeddings}

To illustrate the use of embeddings, we begin with a simple setup in which the outcome function is homogeneous among users, i.e., $f(z,x) = f(z)$ is only a function of the treatments, not the user covariates; we relax this below.
In a typical A/B testing framework, $z$ is a discrete set, which makes estimation difficult when there are a large number of treatments. 
Existing approaches to improve efficiency  using shrinkage~\citep{dimmery2019shrinkage,ignatiadis2019covariate} help alleviate this but still face fundamental limits because of the discrete treatment space. 
However, in our setting, each generated hypothesis (or treatment) is typically associated with an embedding that encodes important information. In the generative setting, we can easily obtain such an embedding by, e.g., using the last layer of a deep neural network that generated the treatment.
By representing the treatment space $\bZ$ as an embedding space, with each treatment lying in $\mathbb{R}^p$ (possibly with large $p$), instead of a space of unrelated discrete elements, we avoid some of the difficulties associated with high cardinality treatments by implicitly pooling observations across treatments that are close in embedding space. 


While conceptually straightforward, the continuous parameterization raises several estimation challenges. We tackle those using kernel methods, which is easily adapted to measuring similarity between treatments over the embedding space $\cZ$. In particular, we apply kernel ridge regression (based on a kernel $\cK$) to obtain an estimator $\hat{f}$ for the outcome response curve. Given sample $(z_i, y_i)$, we fit the following penalized regression: 
\begin{align}
    \hat f = 
    \argmin_{f\in \bbH_\cK} \frac{1}{2n}\sum_{i=1}^n
    (y_i - f(z_i))^2 
    + 
    \lambda_n \|f\|_{\bbH_\cK}.
\end{align}
We then predict the estimated outcome response given a specific treatment $z\in \cZ$ as $\hat{f}(z)$. See Appendix~\ref{app:kernel_methods} for additional details.

\subsection{Warmup: Incorporating covariates}\label{sec:model}

Building on this simple case, we can then extend the idea of incorporating treatment embeddings into a general setup that also includes user covariates. In an idealized setting, we would estimate the kernels with access to all $n\times n$ (expected) potential outcomes $\{f(z_k, x_i)\}_{k,i\in[n]}$,
For example, a natural approach for this structure is the multi-task Gaussian process framework introduced by \citet{bonilla2007multi}, which defines $\Kg$ and $ \Kh$ as kernel functions that depict the similarity between treatments and individual covariates. In practice, however, the fundamental problem of causal inference makes directly estimating $f(z,x)$ challenging because most values of this function are missing. In the next section, we propose a kernel-based approach that makes progress despite these hurdles.

\section{Learning with treatment embeddings and individual covariates}\label{sec:methodology}

We now present our main approach, which we call \emph{Double Kernel Representation Learning} for treatment effect estimation. 
In section~\ref{sec:dkl}, we introduce a factorization that captures the estimation problem as an inner product of representations involving treatment embeddings and user attributes. We describe an alternating minimization style algorithm that recovers these representations. In Section~\ref{sec:analysis}, we provide convergence guarantees of our estimator constructed from these representations under a fixed basis.


\subsection{Double-kernel representation learning}\label{sec:dkl}

\subsubsection{Setup and problem formulation}

For binary treatment, \citet{nie2021quasi} considered the following reformulated outcome model:
\begin{align}\label{eqn:plm}
    y = f(z_0, x) + \bone\{z = z_1\} \cdot \tau(x). 
\end{align}
Define the outcome model $m(x) = \E{y\mid x}$ and propensity score $e(x) = \E{\bone\{z = z_1\}\mid x}$. 
\citet{nie2021quasi} introduced the Robinson Decomposition \citep{robinson1988root} and further transformed \eqref{eqn:plm} into the following partial linear model:
\begin{align}
    y = m(x) + (\bone\{z = z_1\} - e(x)) \cdot \tau(x). 
\end{align}
\citet{kaddour2021causal} extended this approach to the setting where $z$ is a high-dimensional or continuous treatment by directly assuming $f(z,x)$ in \eqref{eqn:fzx} can be factorized as follows:
\begin{align}
    f(z,x) = \bg(z)^\top \bh(x), 
\end{align}
where $\bg: \bbR^p \to \bbR^r$ and $\bh: \bbR^q \to \bbR^r$ are mappings in the real vector space.

Here, we take a slightly different perspective: instead of factorizing $f(z,x)$, we factorize the treatment effect $\tau(z,x)$, similar to \citet{nie2021quasi}. We posit the following partial linear model for $f$:
\begin{align}
    f(z, x) 
    - 
    f(z_0, x)
    =  
    \bg(z)^\top \bh(x),
\end{align}
which yields:
\begin{align}\label{eqn:y-model}
    y = f(z_0, x) + \bg(z)^\top\bh(x) + \epsilon.  
\end{align}
The above factorization states that the CATE function has a low-dimensional representation despite the high-dimensional covariate and treatment inputs. Such a structure is common in the literature (see Section \ref{sec:related-work}) and has become foundational for estimation and optimization. Following \cite{wainwright2019high, rohde2011estimation}, we can further relax this to be approximately low-dimensional, allowing a sparse set of large signals along with many small signals. 

We can further condition on the user-level covariates by 
reparametrizing the model \eqref{eqn:y-model} and taking the conditional expectation with respect to $x$. Formally, assume the positivity assumption:
\begin{align}
    \Prob{z\mid x} > 0, \text{ for all } z\in\cZ, x\in\cX. 
\end{align}
Let $\overline{\bg}(x) =  \E{\bg(z)\mid x}^\top$. Then, 
\begin{align}\label{eqn:y-model-x}
    m(x) = f(z_0, x) + \overline{\bg}(x)^\top \bh(x).
\end{align}
Taking the difference between \eqref{eqn:y-model} and \eqref{eqn:y-model-x} yields:
\begin{align}\label{eqn:y-decomp-R}
    y = m(x) + (\bg(z) - \overline{\bg}(x))^\top \bh(x) + \epsilon. 
\end{align}
In the special case where randomization to treatment $z$ is independent of $x$, $\overline{\bg}(x) = \overline{\bg}$ is a constant function; we start with this case and still denote $\bg(z) - \overline{\bg}$ as $\bg(z)$. In addition, we can directly use the residuals $y - m(x)$ for estimation by plugging in an estimated outcome model $m(x)$. For simplicity, we therefore take $y$ to be the residualized outcome here without loss of generality.

\subsubsection{Estimation and main result}
Building on the above, a natural idea is to solve for $\bg$ and $\bh$ via the following double kernel regression:
\begin{align}\label{eqn:dkl}
    (\hat{\bg}, \hat{\bh}) =& \arg \min_{\{g_l,h_l\}_{l=1}^r}\frac{1}{2n}\sum_{i=1}^n \lt\{y_i - \sum_{l=1}^r g_l(z_i) h_l(x_i)\rt\}^2
    + \lambda_n \sum_{l=1}^r (\|g_l\|_{\Kg}^2 + \|h_l\|_{\Kh}^2).
\end{align}
We then prove the following representer theorem (see Appendix~\ref{app:kernel_methods} and ~\ref{app:repr_thm_proof}):
\begin{theorem}[Representer theorem]
\label{thm:representor}
    The optimal solutions $\bg^\star$ and $\bh^\star$ to the optimization problem~\ref{eqn:dkl} lie in $\bbH_{\Kg}^S$ and $\bbH_{\Kh}^S$ respectively.
\end{theorem}
The representer theorem immediately suggests that the solutions to $\hat \bg$ and $\hat \bh$ have the following form:
\begin{align}\label{eqn:g-h-rkhs}
   \hat \bg(z) = \sum_{i\in[n]} \bUr_i \Kg(z, z_i), 
    \quad 
    \hat \bh(x) = \sum_{i\in[n]} \bVr_i \Kh(x, x_i),
\end{align}
where the vectors $\bUr_{i}, \bVr_i \in \bbR^r$. For convenience, we capture them using the following matrices: $\bU = [\bUr_1, \dots, \bUr_n]^\top \in \bbR^{n\times r}$ and $\bV = [\bVr_1, \dots, \bVr_n]^\top \in \bbR^{n\times r}$, 
each row representing a feature representation for a unit. Alternatively, we can write $\bU$ and $\bV$ in terms of their columns: $\bU = [\bUc_1, \dots, \bUc_r]$ and $\bV = [\bVc_1, \dots, \bVc_r]$.
With \eqref{eqn:g-h-rkhs}, we have a double RKHS model for $f$:
\begin{align}
    f(x,t) = &\sum_{i\in[n]}\sum_{j\in[n]}\innerprod{\bUr_i}{\bVr_j}\Kg(z, z_i)\cK_h(x, x_i)
    =
    \Kg(z, \bz_{1:n})\bTheta\cK_h(x, \bx_{1:n})^\top,
\end{align}
where $\bTheta = \bU\bV^\top \in \bbR^{n\times n}$ is a low-rank matrix. We  therefore propose the following double RKHS regression:
{\small
\begin{align}\label{eqn:rkhs-uv}
     (\hat\bU, \hat\bV) 
    & = \argmin_{\bU,\bV} \frac{1}{2n}\sum_{k=1}^n \{y_k - \Kg(z_k, \bz_{1:n}) \bU\bV^\top \Kh(x_k, \bx_{1:n})^\top\}^2 
    +
    \lambda_n \sum_{l\in[r]} ({\bUc_l}^\top \Kg \bUc_l + {\bVc_l}^\top \Kh \bVc_l).
\end{align}
}

Computationally, \eqref{eqn:rkhs-uv} can be solved by alternatively minimizing over $\bU$ and $\bV$. Concretely, suppose we have $(\hat{\bU}^{(t)}, \hat{\bV}^{(t)})$ from the $t$-th iteration. In the $(t+1)$-th round, we update the parameters as follows: 
{\small 
\begin{align}
    \hat\bU^{(t+1)} 
    = & 
    \argmin_{\bU} \frac{1}{2n}\sum_{k=1}^n \{y_k - \Kg(z_k, \bz_{1:n}) \bU\bV^{(t)\top} \Kh(x_k, \bx_{1:n})^\top\}^2 
     +
    \lambda_n \sum_{l\in[r]} {\bUc_l}^\top \Kg \bUc_l, \label{eqn:U-step}\\
    \hat\bV^{(t+1)} 
    = & 
    \argmin_{\bV} \frac{1}{2n}\sum_{k=1}^n \{y_k - \Kg(z_k, \bz_{1:n}) \bU^{(t+1)}\bV^{\top} \Kh(x_k, \bx_{1:n})^\top\}^2 
     +
    \lambda_n \sum_{l\in[r]} {\bVc_l}^\top \Kh \bVc_l. \label{eqn:V-step}
\end{align}
}
Both \eqref{eqn:U-step} and \eqref{eqn:V-step} have closed-form solutions when viewed as weighted ridge regression. Moreover, the alternating minimization process guarantees a descending (thus convergent) loss function:
\begin{align}
    L(\hat\bU^{(t)}, \hat\bV^{(t)})
    \ge
    L(\hat\bU^{(t+1)}, \hat\bV^{(t)})
    \ge
    L(\hat\bU^{(t+1)}, \hat\bV^{(t+1)}). 
\end{align}
We summarize the procedure in Algorithm \ref{alg:UV}\footnote{There are other methods in the literature that may be used for solving the program, such as (stochastic) gradient descent \citep{zhou2012kernelized} and Quasi-Newton methods \citep{abernethy2006low}.}, along with some discussions.  


\begin{algorithm}[ht!]
   \caption{Double Kernel Learning via Alternating Projection}
   \label{alg:UV} 
{\small 
\begin{algorithmic}
   \STATE {\bfseries Input:} data $(x_i, z_i, y_i)$, kernel $\cK_g, \cK_h$, rank $r$, penalty level $\lambda_n$, maximal iteration $T$, accuracy $\texttt{tol}$
   \STATE Initialize $\bU^{(0)}$ and $\bV^{(0)}$ as random matrices with gaussian elements.
   \FOR{$t=0$ {\bfseries to} $T$}
   \STATE Update $\bU^{(t)}\to\bU^{(t+1)}$ by solving \eqref{eqn:U-step};
   \STATE Update $\bV^{(t)}\to\bV^{(t+1)}$ by solving \eqref{eqn:V-step}; 
   \IF{$\|\bU^{(t)}-\bU^{(t+1)}\|_F/\|\bU^{(t)}\|_F<\texttt{tol}$ 
   and
   $\|\bV^{(t)}-\bV^{(t+1)}\|_F/\|\bV^{(t)}\|_F<\texttt{tol}$}
   \STATE Break
   \ENDIF
   \ENDFOR
   \STATE \textbf{Output: } $\bU^{(t)}$, $\bV^{(t)}$
\end{algorithmic}
}
\end{algorithm}

\textit{Complexity of Algorithm \ref{alg:UV}.} Algorithm \ref{alg:UV} involves a loop that alternates between solving  \eqref{eqn:U-step} and \eqref{eqn:V-step}, thus the computational complexity depends on the solver for  \eqref{eqn:U-step} and \eqref{eqn:V-step}. In our implementation, we solve  \eqref{eqn:U-step} and \eqref{eqn:V-step} by updating the columns of $\bU$ and $\bV$ recursively via ridge regression with a general quadratic form penalty. To run these ridge regressions, we can compute the inverse of the kernel matrices before the loop. With these holdout inverse matrices, we only need $O(N^2)$ to solve each round of the generalized ridge regression. Therefore, in total, the time complexity is
\begin{align*}
    O(\underbrace{N^3}_{\text{Compute inversion of kernel matrices}}) 
    \quad + \quad 
    O(\underbrace{T}_{\text{num of loops}} \cdot \underbrace{r}_{\text{Rounds of ridge regressions}} \cdot \underbrace{N^2}_{\text{compute ridge updates}}).
\end{align*}

\textit{Scalability to large-scale datasets.} 
To make the algorithm adaptive to large-scale datasets (especially with large $N$), we can leverage standard techniques to speed up kernel computation. For example, using the Nystr\"{o}m method \citep{williams2000using} and its variations can reduce the inversion complexity to linear in $N$. 

\textit{Convergence of the algorithm.} The optimization performance of alternative minimization is well studied. Since we are more  concerned with the statistical properties (see Theorem \ref{thm:hGamma-bound}) and experimentation performance (see Theorem \ref{thm:regret}) of the proposed method, we refer interested readers to optimization-based discussions in \cite{jain2013low, chi2019nonconvex}.

\textit{Comparison with a deep-learning based framework.} Deep-learning offers an alternative framework for learning CATE from \eqref{eqn:y-model} \citep{kaddour2021causal}. That approach focuses primarily on prediction, often resulting in difficult-to-interpret representations for the learned treatment and covariates. The corresponding theoretical results are also quite different: as we show in Section \ref{sec:analysis} below, we provide rigorous theoretical guarantees for the estimation accuracy of reconstructing feature representations. By contrast, theoretical results for deep learning-based approaches mainly focus on proving bounds on excess risk as a metric to measure prediction accuracy, but do not provide results on how well the representations themselves are constructed. Finally, deep learning based approaches are computationally expensive relative to the low computational cost of kernel-based methods.

\textit{Connection with classical approaches. } The proposed double kernel learning method has a deep connection with two classical approaches: (i) kernelized probability matrix factorization, which was introduced by \citet{zhou2012kernelized} and serves as the Gaussian process counterpart of our method; and (ii) low-rank matrix factorization, which explores low-rank signals without a kernel structure and has been discussed extensively in the literature \citep{recht2010guaranteed, chi2019nonconvex}. We explore these connections in more depth in Section \ref{sec:connection} of the Appendix.

\subsection{Theoretical analysis for fixed bases}\label{sec:analysis}
For our theoretical analysis, we follow \citet{kaddour2021causal} and consider a fixed basis setting.
Suppose there are $d_1$ treatment bases $\cZ = \{z^k\}$, which is a set of $p$-dimensional vectors, and $d_2$ user bases $\cX = \{x^k\}$, which is a set of $q$ dimensional vectors. $\cZ$ and $\cX$ summarize the information of the treatment to be tested and the pool of candidate users, respectively. Stack the bases into two matrices:
\begin{align}
    \bZ = [z^1, \dots, z^{d_1}] \in \bbR^{p \times d_1}, 
    \quad 
    \bX = [x^1, \dots, x^{d_2}] \in \bbR^{q\times d_2}. 
\end{align}
Let $(e_z, e_x)$ be a pair of independent uniform samples in the product set $\cZ \times \cX$. The observation $i$ is associated with one covariate $x_i = \bX e_{x,i}$ and one treatment $z_i = \bZ e_{z,i}$. Based on our discussion in Section \ref{sec:dkl}, under the decomposition model \eqref{eqn:y-decomp-R}, there exists a low rank matrix $\bTheta^\star$, such that the observed outcome is 
\begin{align}
    y_i 
    = 
    m(x_i) + z_i^\top \bTheta^\star x_i + \epsilon_i 
    = 
    m(x_i) + e_{z,i}^\top\bZ^\top \bTheta^\star \bX e_{x,i} + \epsilon_i. 
\end{align}
Define the new matrix 
\begin{align}\label{eqn:bGamma-star}
    \bGamma^\star = \bZ^\top \bTheta^\star \bX \in \bbR^{d_1\times d_2}, 
\end{align}
whose rank is at most $r$ if $\text{rank}(\bTheta^\star) \le r$. We prove the following theorem: 
\begin{theorem}\label{thm:hGamma-bound}
    Assume that the noise $\epsilon_i$'s are i.i.d. and sub-exponential, and the true signal matrix $\bGamma^\star$ has rank at most $r$ and entries bounded by $\alpha^\star$. 
    With probability greater than $1 - c_1'\exp(-c_2'd\log d)$ with $d = d_1 + d_2$, 
    the estimator $\hat{\bGamma}$ satisfies 
    \begin{align}
        \frac{1}{d_1d_2}\|\hat{\bGamma} - \bGamma^\star\|_F^2 
        \le C\lambda_n^2 r, 
    \quad \text{where} \quad 
        \lambda_n 
        \asymp 
        \max\lt\{\frac{1}{\sqrt{n}}\|\hat{m}(x) - m(x)\|_2, \sqrt{\frac{d\log d}{n}}\rt\}. 
    \end{align}
\end{theorem}

Theorem \ref{thm:hGamma-bound} suggests that $\hat{\bGamma}$ is consistent in large samples as long as $\lambda_n^2 r \to 0$. When the rank $r$ is fixed, consistency then only requires a vanishing penalization level $\lambda_n$. 
We can then consider each element of $\lambda_n$ in turn. 
The left term captures the error in the outcome regression function, $m(x)$. For this term to vanish, we need a consistent estimator $\hat{m}(x)$ of $m(x)$ in the sense that $({1}/{\sqrt{n}})\|\hat{m}(x) - m(x)\|_2 \to 0$. This is guaranteed by many statistical/machine learning methods $\hat{m}$ under the assumption that $m(x)$ belongs to an appropriate function class. For example, if $\hat{m}(x)$ is fit with LASSO, the loss vanishes at a rate of $\sqrt{s \log p /n}$ where $p$ is the dimension of the features and $s$ is the sparsity level \citep{bickel2009simultaneous}. If $\hat{m}(x)$ is fit via RKHS regression with a Gaussian kernel, the rate is $\sqrt{{\log n}/{n}}$ \citep{wainwright2019high, ma2023optimally}. 
The right term, $\sqrt{d\log d/n}$, captures the cost of low-rank matrix recovery. For this term to vanish, $n \gg dr\log d $, where $dr$ captures the intrinsic dimension of a low-rank matrix \citep{candes2011tight}. This rate has been verified to be minimax-optimal for matrix completion tasks \citep{rohde2011estimation, negahban2012restricted, klopp2014noisy}. Finally, following \cite{wainwright2019high, rohde2011estimation}, we note that it is possible to relax this setup to be approximately low-rank, in the sense that the true signal can admit a set of large signals along with many small signals. We relegate a comprehensive analysis under the approximate low-rank regime to Appendix \ref{sec:proof}.


\section{Online Experimentation: An Adaptive Strategy}\label{sec:etc}
In this section, we provide an adaptive algorithm for assigning treatments to individuals as they become available in an online manner. We use Theorem \ref{thm:hGamma-bound} to justify the performance of an explore-then-commit strategy. Consider a bandit with $T$ rounds, which we divide into two stages. In the first stage, we use $T_e$ rounds to explore the best arm for each user by drawing some samples and learning the outcome response matrix $\Gamma$. In the second stage, we keep pulling from the learned best arm for each selected user. Algorithm \ref{alg:etc} summarizes this approach. 

\begin{algorithm}[tb]
   \caption{Explore-then-commit}
   \label{alg:etc} 
{\small    
\begin{algorithmic}
   \STATE {\bfseries Input:} Exploration rounds $T_e$ and exploitation rounds $T_c$, total rounds $T$, treatment bases $\cZ$, covariates $\cX$
   \STATE \texttt{\# Data collection for exploration:}
   \FOR{$t=1$ {\bfseries to} $T_e$} 
   \STATE Randomly sample $e_{z,t}$ with replacement and obtain a treatment $z_t = e_{z,t}^\top \bZ$;
   \STATE Randomly sample a $e_{x,t}$ with replacement and obtain a target user with covariate $x_t = e_{x,t}^\top \bX$; 
   \STATE Assign the user to treatment $z_t$ and observe outcome response $y_t$; 
   \ENDFOR
   \STATE \texttt{\# Learn outcome responses:}
   \STATE Learn conditional mean $\hatm(x)$ with machine learning algorithms;
   \STATE Learn a outcome response matrix $\hat{\bGamma}$ by solving \eqref{eqn:dkl};
   \STATE \texttt{\# Commit to the best personalized treatment:}
   \FOR{$t=T_e + 1$ {\bfseries to} $T$}
   \STATE Randomly sample $e_{x,t}$ with replacement from $\cX$; 
   \STATE Assign the user to treatment $z_t$ which maximizes the estimated outcome response vector $\hat{\bGamma}e_{x,t}$;
   \STATE Collect outcome response $y_t$; 
   \ENDFOR
   \STATE {\bfseries Output:} outcome response matrix $\hat\bGamma$, data $(z_t, x_t, y_t)$.
\end{algorithmic}
}
\end{algorithm}

We prove that the explore-then-commit algorithm will lead to a sublinear regret. To see this, we establish the following theorem:
\begin{theorem}\label{thm:regret}
    Assume the conditions in Theorem \ref{thm:hGamma-bound}. Also, assume that $m(x)$ lies in the RKHS generated by a Gaussian kernel. 
    The explore-then-commit algorithm has the following sublinear regret bound with probability greater than $1 - c_1'\exp(-c_2'd\log d)$:
    $\textup{Regret} 
        \le 
        CT^{2/3}d\log d^{1/3}r^{1/3}, 
    $ where $d = d_1 + d_2$. 
\end{theorem}
We can see that the simple Explore-then-Commit strategy suffices to guarantee a regret with rate $O(T^{2/3})$, which is sublinear. We anticipate that we can use ideas from the literature on low-rank bandits, such as \citet{jun2019bilinear} and \citet{lu2021low}, to sharpen this regret bound in terms of the horizon $T$, though this is outside the scope of the current paper.

\section{Experimental Evaluation}\label{sec:simulation}

In this section, we conduct semi-synthetic experiments based on three open-source datasets. (i) The Upworthy Research Archive \citep{matias2021upworthy} is an open dataset of thousands of A/B tests of headlines conducted by Upworthy from January 2013 to April 2015. We use the Exploratory Dataset in The Upworthy Research Archive, which contains headlines and metrics (clicks, impressions, click-through rates, or CTR) from thousands of experiments. (ii) The MIND dataset \citep{wu2020mind} is a benchmark dataset containing traffic from Microsoft for click rate prediction and news recommendation.  (iii) The ASOS E-Commerce Dataset\footnote{The data is available at \url{https://huggingface.co/datasets/TrainingDataPro/asos-e-commerce-dataset}} is an open-source A/B testing dataset from ASOS in the fashion industry. Due to space constraints, we present a subset of results regarding the Upworthy experiment and MIND dataset, and relegate the remaining results to Section \ref{sec:more-simulation} in the Appendix. The code for the algorithm and replication of the numerical experiments can be found here: \url{https://github.com/LeiShi-rocks/DKRL-LLM}.

\subsection{The Upworthy experiment}

\textbf{Upworthy experimental setup.} We choose $|\cZ| = 50$ headlines as candidate treatments in a hypothetical experiment. We use the sentence transformer \texttt{MiniLM}~\citep{wang2020minilm} to encode the sampled headlines into sentence embeddings of dimension $p = 384$. Since user-level covariates $x$ are not available in the Upworthy Dataset, we simulate $n = 500$ Gaussian vectors of dimension $q = 200$ as baseline covariates. The outcome of interest is the potential revenue that each user can contribute. When a user with covariate $x$ views the headline $z$, we model the average (centered) potential revenue generated by this particular user as $f(z,x) = z^\top \bTheta^\star x + \epsilon$, 
where $\bTheta^\star\in\bbR^{p\times q}$ is a matrix with varying ranks and $\epsilon$ is additive noise.

\paragraph{Evaluation of the estimation error.} 
We use Monte Carlo simulations to evaluate the train and test error of the proposed method, double kernel representation learning (\texttt{DKRL}), and compare with two baseline methods: (i) Structured Intervention Network (\texttt{SIN}) by \cite{kaddour2021causal}, a deep-learning based framework based on a generalized Robinson decomposition; and (ii) RKHS regression with a product kernel $\cK_g \odot \cK_h$ (\texttt{ProdKernel}). We split the synthetic dataset into a training set and test set, fit different methods on the training data, and evaluate the fitted results on the test dataset; we repeat this procedure across multiple ranks.

\begin{table}[ht]
\centering
\caption{Upworthy Metrics by Rank and Method (Mean (Std), rounded to four decimals)}
\label{tab:upworthy-tab}
{\small 
\begin{tabular}{llccc}
\toprule
\textbf{Rank} & \textbf{Method}         & \textbf{Train Error}         & \textbf{Test Error}          & \textbf{Time (sec)}                \\
\midrule
\multirow{3}{*}{2} 
  & \texttt{SIN}            & 0.0105 (0.0006)     & 0.0104 (0.0017)     & 8.4686 (0.6584)     \\
  & \texttt{DKRL}           & \textbf{0.0019 (0.0004)}     & \textbf{0.0036 (0.0010)}     & 0.1907 (0.0339)     \\
  & \texttt{ProdKernel} & 0.0043 (0.0002)     & 0.0080 (0.0013)     & 0.0014 (0.0002)     \\
\midrule
\multirow{3}{*}{3} 
  & \texttt{SIN}            & 0.0113 (0.0006)     & 0.0113 (0.0016)     & 7.8103 (0.7073)     \\
  & \texttt{DKRL}           & \textbf{0.0022 (0.0002)}     & \textbf{0.0048 (0.0013)}     & 0.5062 (0.1448)     \\
  & \texttt{ProdKernel} & 0.0048 (0.0002)     & 0.0090 (0.0013)     & 0.0014 (0.0002)     \\
\midrule
\multirow{3}{*}{5} 
  & \texttt{SIN}            & 0.0143 (0.0006)     & 0.0144 (0.0018)     & 8.5115 (0.5318)     \\
  & \texttt{DKRL}           & \textbf{0.0025 (0.0001)}     & \textbf{0.0078 (0.0012)}     & 1.6933 (0.3752)     \\
  & \texttt{ProdKernel} & 0.0062 (0.0002)     & 0.0115 (0.0014)     & 0.0015 (0.0003)     \\
\midrule
\multirow{3}{*}{7} 
  & \texttt{SIN}           & 0.0171 (0.0007)     & 0.0172 (0.0021)     & 7.8928 (0.3724)     \\
  & \texttt{DKRL}           & \textbf{0.0025 (0.0001)}     & \textbf{0.0108 (0.0015)}     & 3.6143 (0.5942)     \\
  & \texttt{ProdKernel} & 0.0072 (0.0002)     & 0.0136 (0.0017)     & 0.0015 (0.0005)     \\
\bottomrule
\end{tabular}
}
\end{table}
From Table \ref{tab:upworthy-tab}, we can see that in these cases, DKRL achieves the best training and testing error.
Neural network-based approaches are likely inferior here because they may not learn the low-rank representation; moreover, it is quite challenging to construct the relevant propensity functions for high-dimensional treatments (although this can be avoided in experimental settings). The low computational cost (measured on a Macbook Pro with a M2 Max chip) of kernel-based methods also enables more rapid iteration for large-scale A/B testing.

\begin{figure}[ht!]
\centering
\subfigure[]{\includegraphics[width=0.38\textwidth]{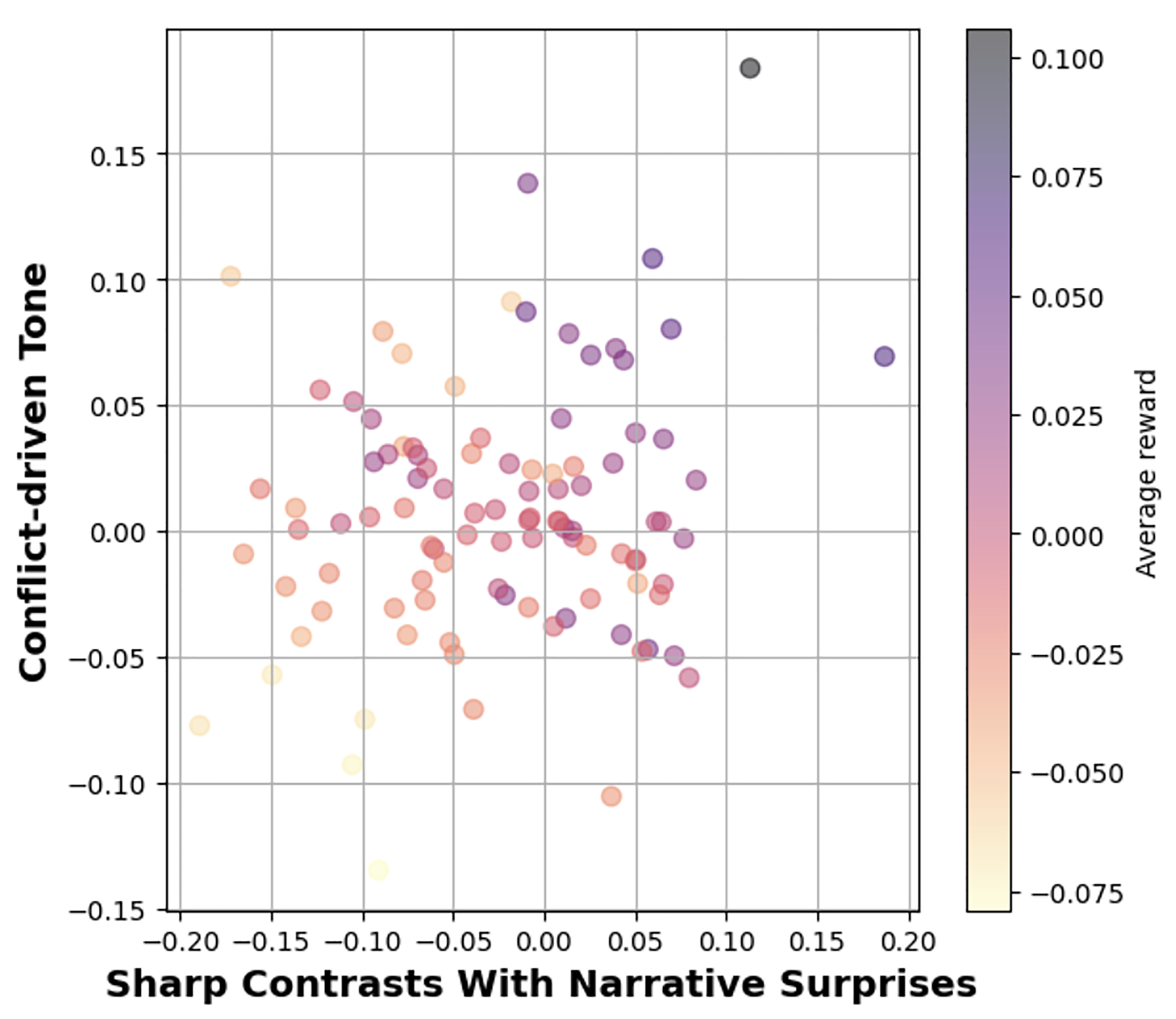}}
\subfigure[]{\includegraphics[width=0.61\textwidth]{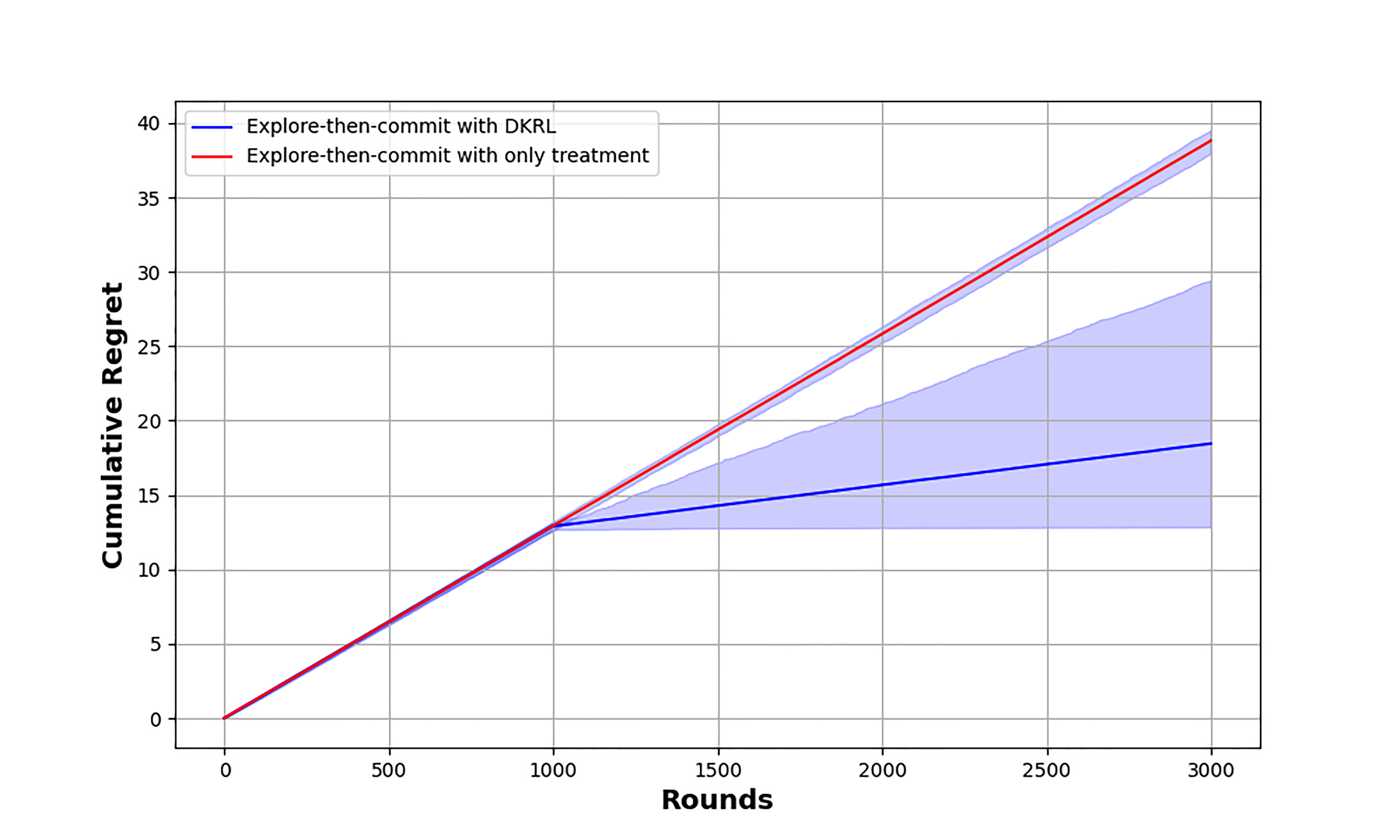}}
\caption{Results on the Upworthy experiment. Panel (a) shows kernel representation of treatments in the semi-synthetic Upworthy Dataset. Panel (b) shows the cumulative regret in the evaluation of the bandit algorithm. The shaded area is the 15\%-85\% quantile range of the regret over simulations. }
\label{fig:upworthy}
\end{figure}

\paragraph{Dimension reduction and interpretation.} Double kernel representation learning also enables low-dimensional summaries of high-dimensional features. Figure \ref{fig:upworthy}(a) presents the kernel representation for the treatment learned from the data. It shows that the components have a semantic interpretation (based on an evaluation by GPT 4o). More concretely, the first learned feature captures the scale of sharp contrasts with narrative surprises; the second captures the extent of a conflict-driven tone. Hence, we can view the values as semantic scores that can directly drive the average reward of the headlines and improve overall interpretability. In Table \ref{tab:upworthy-interpretability} of the Appendix, we show several example headlines with extreme semantic scores, showcasing how the features capture semantic meaning.


\paragraph{Evaluation of the bandit algorithm.} We also evaluate the Explore-then-commit algorithm for online experimentation (Algorithm \ref{alg:etc}). 
We add explore-then-commit with only treatment representations as an additional baseline online experimentation method for comparison.
Figure \ref{fig:upworthy}(b) presents the cumulative regret of Algorithm \ref{alg:etc}. As demonstrated by the theoretical results in Theorem \ref{thm:regret}, Algorithm \ref{alg:etc} produces a sublinear regret in the long run; ignoring user-level covariates leads to failed exploration and thus linear regret. This highlights the importance of incorporating treatment-user interactions into both the outcome modeling and adaptive design strategy.

\subsection{MIND dataset}
Finally, we present results on the MIND dataset, an observational dataset with news recommended to users with unknown probabilities.
We consider a semi-synthetic setup: suppose our goal is to train this recommendation system from scratch by adaptively collecting recommendation-click data to improve recommendation quality. We take $x$ to be user features such as news category preference and historical news click embeddings, and $z$ to be the embeddings for a set of candidate news to be recommended to users. All embeddings are taken from the knowledge graph embeddings that were originally included in the dataset. We consider a synthetic outcome model for the click-through rate: $y = z^\top \bTheta^\star x + \epsilon$, where $\bTheta^\star$ is a matrix that encodes the interaction mechanism between users and treatments; we can interpret this as a match between user preferences and news information. We consider $\bTheta^\star = \bU \bLambda \bV^\top $, with $\bLambda_i = \diag{i^{-q}}$, and vary $q$ to ensure that the interaction model varies from low rank (high eigenvalue decay rates) to  high rank (low eigenvalue decay rates) setting. We compare our method with four baselines: LASSO, XG-boost, a feed-forward neural network, and a kernel regression. We finally combine the methods  with different estimation strategies: ``Z'' only incorporates treatment information, ``X'' only includes covariate information, and ``ZX'' includes a concatenated $(Z,X)$ vector. 

Table \ref{tab:RMSE-MIND} in the Appendix shows test RMSE. Overall, the performance of all methods is better with low-rank signals (larger $q$) than high-rank signals. However, DKRL exploits this low-rank structure most effectively, with the strongest performance. Even with a high-rank interaction matrix (e.g., a full-rank matrix), DKRL effectively exploits the similarity structure in the embeddings, improving accuracy and achieving comparable performance to XG Boost with combined features.

\section{Conclusion}\label{sec:conclusion}
\textit{Summary.} As content creation at scale becomes more readily available, developing new causal inference methods becomes increasingly important. In this work, we explore the design and analysis of GenAI-powered digital experimentation. We propose to use GenAI for efficient and novel treatment generation and to learn the corresponding heterogeneous treatment effect function using a double-kernel representation learning framework. We then propose an adaptive data collection algorithm for online experimentation. Finally, we provide theoretical guarantees and multiple numerical demonstrations. 

\textit{Limitations and future directions.} Several limitations remain.
First, it is important to consider more complex experimentation settings that are important in practice, such as delayed responses \citep{shi2023statistical, shi2024using} or interference between participants \citep{li2022network}. Second, we established theoretical guarantees for the double-kernel representation learning framework in a fixed-basis setting. It would be interesting to generalize this theory to infinite-dimensional bases. Third, GenAI-powered experimentation raises key questions on issues like privacy and fairness, which are important to address independent of this current research. Fourth, it is natural to use other methods such as language model finetuning \citep{hu2022lora} or contrast learning \citep{zha2023rank} to learn improved embeddings and to generally improve the pipeline.

    


\begin{ack}
We thank the AC and four reviewers for constructive comments and feedback.
\end{ack}

{
\small

\bibliographystyle{apalike}
\bibliography{ref}


}


\appendix

\section{Kernel Methods}\label{app:kernel_methods}
Kernel-based methods have been widely used in related practices such as representation learning \citep{esser2024non}, dimension reduction \citep{scholkopf1997kernel}, natural language processing \cite{joachims1998text}, among others. 
We do a brief overview of RKHS methods. Consider a positive definite kernel $\cK: \cZ \times \cZ \to  \bbR$ supported on a compact $d$-dimensional set $\cZ \subset \bbR^d$. A Hilbert space $\bbH_\cK$ of functions on $\cZ$ equipped with an inner product $\innerprod{\cdot}{\cdot}_{\bbH_\cK}$ is called a reproducing kernel Hilbert space (RKHS) with reproducing kernel $K$ if the following properties are satisfied: 
\begin{enumerate}
    \item for all $z \in \cZ$, $\cK(\cdot, z) \in \bbH_\cK$;
    \item for all $z \in \cZ$ and $f \in \bbH_\cK$, $\innerprod{f}{\cK(\cdot, z)}_{\bbH_\cK} = f(z)$  (reproducing property).
\end{enumerate}

A natural idea for estimation and policy learning is to incorporate embedding information with RKHS methods. Given sample $(t_i, y_i)$, we can fit the following penalized nonlinear least squares: 
\begin{align}
    \hat f = 
    \argmin_{f\in \bbH_\cK} \frac{1}{2n}\sum_{i=1}^n
    (y_i - f(z_i))^2 
    + 
    \lambda_n \|f\|_{\bbH_\cK}.
\end{align}
Then we can predict the estimated outcome response given a specific treatment $z\in \cZ$ as $\hat{f}(z)$. By the representer theorem, this leads to a program based on the data:
\begin{align}
    {\halpha}
    =
    \argmin_{\alpha\in \bbR^n} \frac{1}{2n}\sum_{i=1}^n
    \{y_i - \sum_{i=1}^n \alpha_j \cK(z_i, z_j)\}^2 
    + 
    \lambda_n \alpha^\top \cK \alpha,
\end{align}
which gives the interpolator
\begin{align}
    \hat{f}(z) = \sum_{i=1}^n \halpha_j \cK(z, z_j). 
\end{align}
Another viewpoint of the RKHS-based method is that we can put a distribution on the functions and fit a Gaussian process model \citep{williams2006gaussian}: 
\begin{align}\label{eqn:gp}
    y \sim \cN(0, \sigma^2), 
    \quad 
    f\sim \text{GP}(0, \cK). 
\end{align}
Then the interpolator is equivalent to a posterior prediction from \eqref{eqn:gp}.

In particular, if the kernel $\cK$ is a linear inner product of finite dimensional features $\phi(z)$, the above program is also equivalent to ridge regression:
\begin{align}
    \hat{f}(z) = \phi(z)^\top \hbeta,
\end{align}
where
\begin{align}
    \hbeta = \frac{1}{2n}\sum_{i=1}^n \{y_i - \phi(z_i)^\top \beta\}^2 
    +
    \lambda_n\|\beta\|_2^2.
\end{align}
Such connection has been pointed out by previous literature, e.g. \citet{valko2013finite}.

This is also equivalent to a Bayesian linear model:
\begin{align}
    y \sim \cN(\phi(z)^\top\beta, \sigma^2 I), \text{ where } 
    \beta\sim \cN(0, \sigma_0^2 I). 
\end{align}

\section{Connection with classical approaches}\label{sec:connection}

For intuition we now draw connections between the proposed method to two classical approaches: kernelized probability matrix factorization, which is introduced by \citet{zhou2012kernelized} and serves as the Gaussian process counterpart of our method, and low-rank matrix factorization algorithms. 

\paragraph{Connection with kernelized probability matrix factorization.} The proposed method is closely related to kernelized probability matrix factorization \citep{zhou2012kernelized}. To see this, we revisit the program \eqref{eqn:rkhs-uv} but instead of using the parametrization $\bU$ and $\bV$ consider the following transformation:
\begin{align}
    \bR = \cK_g \bU \in \bbR^{n\times r}, \quad \bS = \cK_h \bV \in\bbR^{n\times r}. 
\end{align}
Here, $\bR$ and $\bS$ are the intrinsic representations of the treatment and covariate-level information, respectively. Then \eqref{eqn:rkhs-uv} can be written as
\begin{align}
    (\hat{\bR}, \hat{\bS}) 
    =  \argmin_{\bR,\bS} & 
    \frac{1}{2n}\sum_{k=1}^n
    \lt(y_k - {\bRr_k}^\top {\bSr_k} \rt)^2 
    + \lambda_n\sum_{l=1}^r \lt({\bRc_l}^\top  \Kg^{-1} {\bRc_l} + {\bSc_l}^\top\Kh^{-1} {\bSc_l}\rt). 
\end{align}
We can instead follow the kernelized probabilistic factorization in \citet{zhou2012kernelized}. Under the following data-generating process:
\begin{align}
    \bRc_l \sim \textup{GP}(0, \Kg), 
    \quad 
    \bSc_l \sim \textup{GP}(0, \Kh), 
    \quad
    l \in [r], 
\end{align}
and 
\begin{align}
    y_k \sim \cN({\bRr_k}^\top \bSr_k, \sigma^2), 
\end{align}
\citet{zhou2012kernelized} showed that the log-posterior over the latent features $\bR$ and $\bS$ is given by
\begin{align}
    \log p(\bR,\bS\mid \by, \sigma^2)
    =&
    - \frac{1}{2\sigma^2} \sum_{k=1}^n (y_k - {\bRr_k}^\top \bSr_k)^2
    - \frac{1}{2}\sum_{l=1}^r 
    \lt({\bRc_l}^\top  \Kg^{-1} {\bRc_l} + {\bSc_l}^\top\Kh^{-1} {\bSc_l}\rt)\\
    &- n\log \sigma^2 - \frac{r}{2}\{\log(|\Kg|) + \log(|\Kh|)\}
    +
    C, \label{eqn:log-p}
\end{align}
where $C$ is some universal constant that does not depend on $\bR$ and $\bS$. Maximizing the log-posterior \eqref{eqn:log-p} is equivalent to minimizing 
\begin{align}
    L(\bR,\bS) & = \frac{1}{2\sigma^2} \sum_{k=1}^n (y_k - {\bRr_k}^\top \bSr_k)^2 
    + \frac{1}{2}\sum_{l=1}^r 
    \lt({\bRc_l}^\top  \Kg^{-1} {\bRc_l} + {\bSc_l}^\top\Kh^{-1} {\bSc_l}\rt),
\end{align}
which is equivalent to \eqref{eqn:rkhs-uv}.

\paragraph{Connection with low-rank matrix learning with finite-dimensional features.} 
Consider the kernel given by the inner product of some feature mapping $\phi\in\bbR^p$: 
\begin{align}
    \Kg(z, z') = \phi(z)^\top\phi(z'),
\end{align}
with feature matrix $\Phi = [\phi(z_1), \dots, \phi(z_n)]^\top\in\bbR^{n\times p}$.
Then the kernel matrix has the expression:
\begin{align}
    \Kg = \Phi \Phi^\top.  
\end{align}
We prove the following reformulation results:

\begin{proposition}\label{prop:finite-dim}
Suppose the feature matrix $\Phi$ has full (column or row) rank. Then 
\begin{align}\label{eqn:rkhs-tutv}
    (\hat\bU^\star, \hat\bV^\star) = &\argmin_{\bU,\bV} \frac{1}{2n}\sum_{k=1}^n \{y_k - \phi(z_k)^\top \bU^\star {\bV^\star}^\top \psi(x_k)\}^2 
    + 
    \lambda_n (\|\bU^\star\|_F^2 + \|\bV^\star\|_F^2).
\end{align}
\end{proposition}

Now following Proposition \ref{prop:finite-dim}, we have the following equivalent formulation based on nuclear-norm penalization: 
\begin{proposition}\label{prop:nuclear}
    The program \eqref{eqn:rkhs-tutv} is equivalent to the following low-rank learning program: 
    \begin{align}\label{eqn:rkhs-theta}
       \hat{\bTheta} 
       = 
       \argmin_{\bTheta} \frac{1}{2n}\sum_{k=1}^n \{y_k - \phi(z_k)^\top \bTheta \psi(x_k)\}^2 
       + 
       2\lambda_n  \|\bTheta\|_*.
    \end{align}
\end{proposition}
This equivalence connects the matrix factorization approach with nuclear norm penalization for low-rank matrix estimation. The key insight is due to the variational expression of the nuclear norm \citep{recht2010guaranteed}:
\begin{align}
    \|\bTheta\|_* = \min_{\bU^\star,\bV^\star:\bTheta = \bU^\star\bV^{\star\top}}\frac{1}{2}(\|\bU\|_F^2 + \|\bV\|_F^2). 
\end{align}

\section{Additional results}
\subsection{Bounds on $\bTheta$}
The rate in Theorem \ref{thm:hGamma-bound} also implies a rate for $\bTheta$ when the number of tested variants and candidate users is large in the sense that $d_1 > p$ and $d_2 > q$. To see this, 
we can express $\bTheta^\star$ with $\bGamma^\star$, as defined in \eqref{eqn:bGamma-star}:
\begin{align}
    \bTheta^\star = (\bZ\bZ^\top)^{-1} \bZ {\bGamma^\star} \bX^\top (\bX\bX^\top)^{-1}.
\end{align}
This motivates an estimator:
\begin{align}\label{eqn:bTheta}
    \hat{\bTheta} = (\bZ\bZ^\top)^{-1} \bZ \hat{\bGamma} \bX^\top (\bX\bX^\top)^{-1}. 
\end{align}
Then we have the following bound:
\begin{align}
     \|\hat{\bTheta} - \bTheta^\star\|_F^2 
    \le &
    \rho_{\max}\{(\bZ\bZ^\top)^{-1}\}
    \rho_{\max}\{(\bX\bX^\top)^{-1}\}
    \|\hat{\bGamma} - \bGamma^\star\|_F^2\\
    \le & \rho_{\min}\{(\bZ\bZ^\top/d_1)\}^{-1}
    \rho_{\min}\{(\bX\bX^\top/d_2)\}^{-1}
    \cdot \lambda_n^2 r. 
\end{align}
Therefore, as long as the minimal eigenvalue of $\bZ\bZ^\top$ and $\bX\bX^\top$ are not degenerate, the estimator \eqref{eqn:bTheta} is also consistent.  

\section{Technical proofs}\label{sec:proof}
\subsection{Proof of Theorem \ref{thm:representor}}\label{app:repr_thm_proof}
\begin{proof}[Proof of Theorem \ref{thm:representor}]
  The RKHS space $\bbH_g$ and $\bbH_h$ has the following decomposition:
  \begin{align}
      \bbH_g = \bbH_g^S \oplus \bbH_g^\perp, \quad \bbH_h = \bbH_h^S \oplus \bbH_h^\perp,
  \end{align}
  where $\bbH_g^S$ and $\bbH_h^S$ are the orthogonal projection into the sample-generated function subspace. For any $\bg$ and $\bh$, we have the decomposition
  \begin{align}
      \bg = \bg^S + \bg^\perp, \quad \bh = \bh^S + \bh^\perp.
  \end{align}
  Then for any $i$, $\bg(z_i) = \bg^S(z_i) + \bg^\perp(z_i) = \bg^S(z_i)$, $\bh(x_i) = \bh^S(x_i) + \bh^\perp(x_i) = \bh^S(x_i)$.
  Meanwhile, due to the projection, we have a reduction in the RKHS norm:
  \begin{align}
      \|g_l^S\|_{\Kg}^2 \le \|g_l\|_{\Kg}^2,
      \quad
      \|h_l^S\|_{\Kh}^2 \le \|h_l\|_{\Kh}^2.
  \end{align}
  Therefore, by replacing $\bg$ with $\bg^S$ and also $\bh$ with $\bh^S$, the objective function never increases. Hence, the optimal solution must lie in $\bbH_{\Kg}^S$ and $\bbH_{\Kh}^S$.
\end{proof}

\subsection{Proof of Proposition \ref{prop:finite-dim}}

\begin{proof}[Proof of Proposition \ref{prop:finite-dim}]
\textit{Case 1.} When $n \ge p$ and the feature matrix $\Phi \in \bbR^{n \times p}$ has full column rank, we can conclude that the mapping $\bU \mapsto \Phi^\top \bU $ is a surjective mapping from $\bbR^{n \times r}$ to $ \bbR^{p \times r} $. Therefore, \eqref{eqn:rkhs-uv} is equivalent to solving the feature-based optimization program:
\begin{align}
    (\hat\bU^\star, \hat\bV^\star) = \min_{\bU,\bV} \frac{1}{2n}\sum_{k=1}^n \{y_k - \phi(z_k)^\top \bU^\star {\bV^\star}^\top \psi(x_k)\}^2 
    + 
    \lambda_n  (\|\bU^\star\|_F^2 + \|\bV^\star\|_F^2).
\end{align}

\textit{Case 2.} When $n < p$ and the feature matrix $\Phi \in \bbR^{n \times p}$ has full row rank, we start from the program \eqref{eqn:rkhs-tutv}. For any $\bU^\star$, there exists $\bU$, such that 
\begin{align}
    \Phi \Phi^\top \bU = \Phi \bU^\star,
\end{align}
as one can take $ \bU = (\Phi \Phi^\top)^{-1} \Phi \bU^\star $.
Therefore, we can always reparametrize \eqref{eqn:rkhs-tutv} as \eqref{eqn:rkhs-uv}. 

\end{proof}

\subsection{Proof of Proposition \ref{prop:nuclear}}

\begin{proof}[Proof of Proposition \ref{prop:nuclear}]
The nuclear norm $\|\bTheta\|_*$ of a matrix has the following variational representation: 
\begin{align}
    \|\bTheta\|_* = \min_{\bU\bV^\top = \bTheta} \frac{1}{2}(\|\bU\|_F^2 + \|\bV\|_F^2). 
\end{align}

Using this result, we can deduce the equivalence between the program \eqref{eqn:rkhs-tutv} and \eqref{eqn:rkhs-theta}:
\begin{align}
    & \min_{\bTheta} p
    \frac{1}{2n}\sum_{k=1}^n \{y_k - \phi(z_k)^\top \bTheta \psi(x_k)\}^2 
    + 
    2\lambda_n  \|\bTheta\|_* \\
    = & \min_{\bTheta} 
    \frac{1}{2n}\sum_{k=1}^n \{y_k - \phi(z_k)^\top \bTheta \psi(x_k)\}^2 
    + 
    2\lambda_n  
    \lt\{\min_{\bU^\star,\bV^\star:\bU^\star\bV^{\star\top} = \bTheta} \frac{1}{2} (\|\bU^\star\|_F^2 + \|\bV^\star\|_F^2)\rt\} \\
    = & \min_{\bTheta} \min_{\bU^\star,\bV^\star:\bU^\star\bV^{\star\top} = \bTheta}
    \frac{1}{2n}\sum_{k=1}^n \{y_k - \phi(z_k)^\top \bU^\star\bV^{\star\top} \psi(x_k)\}^2 
    +  
    {\lambda_n} (\|\bU^\star\|_F^2 + \|\bV^\star\|_F^2) \\
    = &  \min_{\bU^\star,\bV^\star}
    \frac{1}{2n}\sum_{k=1}^n \{y_k - \phi(z_k)^\top \bU^\star\bV^{\star\top} \psi(x_k)\}^2 
    +  
    {\lambda_n} (\|\bU^\star\|_F^2 + \|\bV^\star\|_F^2). 
\end{align}
\end{proof}

\subsection{Statement and proof of a more general version of Theorem \ref{thm:hGamma-bound}} \label{sec:pf-hGamma-bound}
In this section, we prove a more general version of Theorem \ref{thm:hGamma-bound}, which extends the results to an approximate low-rank setting. More concretely, let $\rho_k(\bGamma)$ be the $k$-th largest singular value of matrix $\bGamma$. We define the following notion of $\ell_q$-ball in the matrix space:
\begin{definition}[$\ell_q$ ball]
    The $\ell_q$ ball with radius $R_q$ is defined as the following set:
    \begin{align}
        \bbB_q(R_q) = \lt\{\bGamma\in\bbR^{d_1\times d_2}: \sum_{k=1}^{\min\{d_1,d_2\}}\rho_k(\bGamma)^q \le R_q\rt\}, \text{ for } q \in[0,1). 
    \end{align}
\end{definition}
When $q = 0$, this definition becomes the exact low-rank condition. When $q > 0$, the definition allows the existence of many small positive singular values, decaying at a proper rate \citep{negahban2012restricted, rohde2011estimation, shi2024low, shi2020noisy, cui2023robust}. 

\begin{theorem}\label{thm:hGamma-bound-approxlowrank}
    Assume that the noise $\epsilon_i$'s are i.i.d. and sub-exponential, and the true signal matrix $\bGamma^\star$ belongs to the $\ell_q$ ball $\bbB_q(R_q)$ and entry bounded by $a$. 
    With probability greater than $1 - c_1'\exp(-c_2'd\log d)$, 
    the estimator $\hat{\bGamma}$ satisfies 
    \begin{align}
        \frac{1}{d_1d_2}\|\hat{\bGamma} - \bGamma^\star\|_F^2 
        \le CR_q\lambda_n^{2-q}, 
    \quad \text{where} \quad 
        \lambda_n 
        \asymp 
        \max\lt\{\frac{1}{\sqrt{n}}\|\hat{m}(x) - m(x)\|_2, \sqrt{\frac{d\log d}{n}}\rt\}. 
    \end{align}
\end{theorem}

\begin{proof}[Proof of Theorem \ref{thm:hGamma-bound}]

For the convenience of analysis, we introduce the sampling operator $\fS:\bbR^{d_1\times d_2} \to \bbR^{n\times 1}$:
\begin{align}
    \fS(\bGamma) 
    = 
    (e_{z,1}^\top\bGamma e_{x,1}, \dots, 
    e_{z,n}^\top\bGamma e_{x,n})^\top. 
\end{align}
$\fS$ is a linear operator. The adjoint operator, $\fS^\star:\bbR^{n\times 1} \to \bbR^{d_1\times d_2}$, is defined as
\begin{align}
    \fS^\star(\bepsilon) 
    =
    \sum_{i=1}^n \epsilon_i e_{x,i}e_{z,i}^\top. 
\end{align}
The operator satisfies
\begin{align}
    \innerprod{\fS(\bGamma)}{\epsilon}
    =
    \innerprod{\bGamma}{\fS^\star(\epsilon)}. 
\end{align}

    \textbf{Step 1. Deriving a basic inequality.} By the minimization program, for the estimated $\hat{\bGamma}$, we have
\begin{align}
    \frac{1}{2n}\|\by - \hatm(\bx) -  \fS(\hat{\bGamma})\|_2^2 + \lambda_n \|\hat{\bGamma}\|_* 
    \le 
    \frac{1}{2n}\|\by - \hatm(\bx) -  \fS(\bGamma^\star)\|_2^2
    + \lambda_n \|\bGamma^\star\|_*. 
\end{align}
This gives
\begin{align}
    & \frac{1}{2n}\|\fS(\hat{\bGamma} - \bGamma^\star)\|_2^2 \\
    & \le 
    \frac{1}{n}\innerprod{\hepsilon}{ \fS(\hat{\bGamma} - \bGamma^\star)} + \lambda_n\|\bGamma^\star\|_* - \lambda_n\|\hat{\bGamma}\|_* \\
    & = \frac{1}{n}\innerprod{\hatm(x) - m(x)}{\fS(\hat{\bGamma} - \bGamma^\star)}
    \\
    & + 
    \frac{1}{n}\innerprod{ \fS^\star(\epsilon)}{\hat{\bGamma} - \bGamma^\star} + \lambda_n\|\bGamma^\star\|_* - \lambda_n\|\hat{\bGamma}\|_*\\
    & \le 
    \frac{1}{n} \|\hatm(x) - m(x)\|_2 \|\fS(\hat{\bGamma} - \bGamma^\star)\|_2
    +
    \frac{1}{n}\| \fS^\star(\epsilon)\|_{\mathrm{op}}\|\hat{\bGamma} - \bGamma^\star\|_* 
    + \lambda_n\|\bGamma^\star\|_* - \lambda_n\|\hat{\bGamma}\|_*.
    \label{eqn:interim-bd}
\end{align}

\textbf{Step 2. A cone inequality for the difference. }
Now we consider the projection of the difference $\hat{\bGamma} - \bGamma^\star$ onto a subspace that will lead to a decomposition into a low-rank part and a remainder that is relatively small in scale.

Let the singular value decomposition (SVD) of $\bGamma^\star$ be $\bGamma^\star = \bU \bLambda \bV^\top$, where $\bU \in \bbR^{d_1 \times d_1}$ and $\bV \in \bbR^{d_2 \times d_2}$ are orthonormal matrices and $\bLambda$ is the singular value matrix with descending diagonals. 
For any matrix $\bDelta \in \bbR^{d_1 \times d_2}$, if we write
\begin{align}
    \bU^\top \bDelta \bV = 
    \begin{pmatrix}
        \bOmega_{11} & \bOmega_{12}\\
        \bOmega_{21} & \bOmega_{22}
    \end{pmatrix}, 
\end{align}
with 
$\bOmega_{11}\in\bbR^{r\times r}$ and $\bOmega_{22}\in\bbR^{(d_1-r)\times(d_2-r)}$. We can define a projection
\begin{align}
    \Pi(\bDelta) = 
    \begin{pmatrix}
        \bOmega_{11} & \bOmega_{12}\\
        \bOmega_{21} & 
        \bzero_{(d_1-r) \times (d_2-r)}
    \end{pmatrix}, 
    \quad 
    \Pi^\perp(\bDelta) = 
    \begin{pmatrix}
        \bzero_{r \times r} & \bzero_{r \times (d_2-r)} \\
        \bzero_{(d_1-r) \times r} & \bOmega_{22}
    \end{pmatrix}. 
\end{align}
With this decomposition, we always have $\textup{rank}(\Pi(\bDelta)) \le 2r$. Now we choose an effective rank by trimming the matrix using a threshold, $\tau$, and we will discuss how to pick $\tau$ later.  

Since $\bGamma^\star\in \bbB_q(R_q)$, we have
\begin{align}\label{eqn:r-inequality}
r\tau^q\le\sum_{k=1}^r\rho_k(\bDelta)^q \le\sum_{k=1}^m \rho_k(\bDelta)^q \le R_q, \quad r\le R_q \tau^{-q}.
\end{align}
Meanwhile, we also have
\begin{align}
\|\Pi^\perp(\bDelta)\|_*=\sum_{k=r+1}^m\rho_k(\bDelta) \le\tau\sum_{k=r+1}^m\big(\frac{\rho_k(\bDelta)}{\tau}\big)^q\le R_q\tau^{1-q}. \label{eqn:Pi-perp-norm}
\end{align}

Using the triangle inequality, we have
\begin{align}
    \|\hat{\bGamma}\|_* 
    & =
    \|\bGamma^\star 
    +
    \hat{\bGamma} 
    - 
    \bGamma^\star\|_*  \\
    & = 
    \|\bGamma^\star 
    +
    \Pi(\hat{\bGamma} 
    - 
    \bGamma^\star)
    +
    \Pi^\perp(\hat{\bGamma} 
    - 
    \bGamma^\star)\|_* \\
    & \ge
    \|\Pi(\bGamma^\star)
    +
    \Pi^\perp(\hat{\bGamma} 
    - 
    \bGamma^\star)
    \|_*
    - 
    \|\Pi(\hat{\bGamma} 
    - 
    \bGamma^\star)\|_* 
    -
    \|\Pi^\perp(\bGamma^\star)\|_*\\ 
    & = 
    \|\Pi(\bGamma^\star)\|_* 
    + 
    \|\Pi^\perp(\hat{\bGamma} 
    - 
    \bGamma^\star)\|_* 
    - 
    \|\Pi(\hat{\bGamma} 
    - 
    \bGamma^\star)\|_*
    -
    \|\Pi^\perp(\bGamma^\star)\|_*, 
\end{align}
which further gives 
\begin{align}
    \|\bGamma^\star\|_* - \|\hat{\bGamma}\|_* 
    \le 
    \|\Pi(\hat{\bGamma} 
    - 
    \bGamma^\star)\|_* -
    \|\Pi^\perp(\hat{\bGamma} 
    - 
    \bGamma^\star)\|_*
    +
    2\|\Pi^\perp(\bGamma^\star)\|_*. 
\end{align}

By choosing $\lambda_n$ that satisfies
\begin{align}
    \lambda_n \ge \max\lt\{\frac{1}{\sqrt{n}}\|\hatm(x) - m(x)\|_2 , \frac{2}{n}\|\fS^\star(\epsilon)\|_\textup{op}\rt\}, 
\end{align}
together with \eqref{eqn:interim-bd}, we have
\begin{align}
    &\frac{1}{2n}\|\fS(\hat{\bGamma} - \bGamma^\star)\|_2^2 + \frac{\lambda_n}{2}\|\Pi^\perp(\hat{\bGamma} - \bGamma^\star)\|_*\\
    \le& 
   \frac{\lambda_n}{\sqrt{n}} \|\fS(\hat{\bGamma} - \bGamma^\star)\|_2
    +
    \frac{3 \lambda_n}{2} \|\Pi(\hat{\bGamma} - \bGamma^\star)\|_*
    +
    2\lambda_n \|\Pi^\perp( \bGamma^\star)\|_*. \label{eqn:interim-bd-2}
\end{align}
For the term $2n^{-1}\|\fS^\star(\epsilon)\|_{\mathrm{op}}$, \citet{negahban2012restricted} established the following bound in the proof of their Corollary 1:
\begin{align}
    \Prob{2n^{-1}\|\fS^\star(\epsilon)\|_{\mathrm{op}}\ge
    c_1\nu \sqrt{d\log d/n} } \le c_2\exp(-c_3d\log d). 
\end{align}

\textbf{Step 3. A final bound.} 
We will prove the following result:

Consider two cases. 

\textit{Case 1. } Suppose we have 
\begin{align}
    \frac{1}{4n}\|\fS(\hat{\bGamma} - \bGamma^\star)\|_2^2 
    \ge 
    \|\hat{\bGamma} - \bGamma^\star\|_F^2. 
\end{align}
Now using
\begin{align}
    \|\Pi(\hat{\bGamma} - \bGamma^\star)\|_* \le \sqrt{2r} \|\hat{\bGamma} - \bGamma^\star\|_F  
\end{align}
and
\begin{align}
    \frac{\lambda_n}{\sqrt{n}} \|\fS(\hat{\bGamma} - \bGamma^\star)\|_2 
    \le
    \lambda_n^2 + \frac{1}{4n}\|\fS(\hat{\bGamma} - \bGamma^\star)\|_2^2, \see{Young's Inequality}
\end{align}
with \eqref{eqn:interim-bd-2}, we have
\begin{align}\label{eqn:interim-bd-3}
    \|\hat{\bGamma} - \bGamma^\star\|_F^2
    \le 
    \frac{1}{4n}\|\fS(\hat{\bGamma} - \bGamma^\star)\|_2^2  
    \le 
    \lambda_n^2
    +
    \frac{9 \lambda_n^2 r}{4} + \frac{1}{2}\|\hat{\bGamma} - \bGamma^\star\|_F^2 + 2\lambda_n\|\Pi^\perp(\bGamma^\star)\|_*.
\end{align}
Using \eqref{eqn:r-inequality} and \eqref{eqn:Pi-perp-norm}, the above gives
\begin{align}
    \|\hat{\bGamma} - \bGamma^\star\|_F^2 \le \frac{26\lambda_n^2 R_q \tau^{-q} }{4} + 4\lambda_n R_q \tau^{1-q}.
\end{align}
To minimize the bound above, the best $\tau$ to pick is $ \tau = 26q(1-q)^{-1}\lambda_n/16 $, which leads to 
\begin{align*}
    \|\hat{\bGamma} - \bGamma^\star\|_F^2 \le CR_q \lambda_n^{2-q}. 
\end{align*}

\textit{Case 2. } Suppose Case 1 fails so that we have
\begin{align}
    \frac{1}{4n}\|\fS(\hat{\bGamma} - \bGamma^\star)\|_2^2 
    \le 
    \|\hat{\bGamma} - \bGamma^\star\|_F^2.
\end{align}
\eqref{eqn:interim-bd-2} gives
\begin{align}
    \frac{\lambda_n}{2}\|\Pi^\perp(\hat{\bGamma} - \bGamma^\star)\|_*
    \le 
   {2\lambda_n} \|\hat{\bGamma} - \bGamma^\star\|_F
    +
    \frac{3 \lambda_n}{2} \|\Pi(\hat{\bGamma} - \bGamma^\star)\|_*
    +
    2\lambda_n\|\Pi^\perp(\bGamma^\star)\|_*,
\end{align}
and from this, we can derive
\begin{align}
    \|\hat{\bGamma} - \bGamma^\star\|_* \le 8\sqrt{2r}\|\hat{\bGamma} - \bGamma^\star\|_F + 4\|\Pi^\perp(\bGamma^\star)\|_*.
\end{align}

\textit{Case 2.1.} Suppose 
\begin{align}
    \|\hat{\bGamma} - \bGamma^\star\|_F^2 
    \le 
    c_0 (\sqrt{d_1d_2}\|\hat{\bGamma} - \bGamma^\star\|_\infty)
    \|\hat{\bGamma} - \bGamma^\star\|_*
    \sqrt{\frac{d\log d}{n}}.
\end{align}
Note that $\lambda_n \ge C \sqrt{{d\log d}/{n}}$ by definition. The equality might not directly hold due to the fitting error for $m(x)$. It is of great interest to see whether such equality can be attained with methods
such as the cross-fitting technique to avoid double-dipping and reduce bias \citep{chernozhukov2018double, lu2025conditional}.

Now Using that 
\begin{align}
    \sqrt{d_1d_2}\|\hat{\bGamma} - \bGamma^\star\|_\infty
    \le {2a}, 
    \quad 
    \|\hat{\bGamma} - \bGamma^\star\|_* \le 
    8\sqrt{2r}\|\hat{\bGamma} - \bGamma^\star\|_F + 4\|\Pi^\perp(\bGamma^\star)\|_*,
\end{align}
and the choice of $\tau = C_q\lambda_n$ in Case 1, we obtain the inequality
\begin{align}
    \|\hat{\bGamma} - \bGamma^\star\|_F^2 \le CR_q^{1/2}\lambda_n^{1-q/2}\|\hat{\bGamma} - \bGamma^\star\|_F + CR_q\lambda_n^{2-q}.
\end{align}
Applying Young's inequality again, we obtain
\begin{align}
    \|\hat{\bGamma} - \bGamma^\star\|_F^2
    \le C R_q\lambda_n^{2-q}.
\end{align}

\textit{Case 2.2. }
Suppose we have that 
\begin{align}
    \frac{384d}{\sqrt{n}}
    \cdot \frac{\|\hat{\bGamma} - \bGamma^\star\|_\infty}{\|\hat{\bGamma} - \bGamma^\star\|_F} > \frac{1}{2}, 
\end{align}
then $\|\hat{\bGamma} - \bGamma^\star\|_F \le {1536 a}/{\sqrt{n}}$ and we obtain the desired rate directly. 

\textit{Case 2.3. } Suppose we are under neither \textit{Case 2.1} and \textit{Case 2.2}. Then the following inequality holds with probability at least $1 - c_1'\exp(-c_2'd\log d)$ by our proof in Step 4:
\begin{align}
    \frac{1}{\sqrt{n}}\|\fS(\hat{\bGamma} - \bGamma^\star)\|_2 
    \ge 
    \frac{1}{8}\|\hat{\bGamma} - \bGamma^\star\|_F 
    -
    \frac{48  d \|\hat{\bGamma} - \bGamma^\star\|_\infty}{\sqrt{n}}
    \ge\frac{1}{16}
    \|\hat{\bGamma} - \bGamma^\star\|_F. 
\end{align}
Now using \eqref{eqn:interim-bd-2}, we obtain that 
\begin{align}
    \frac{1}{512}\|\hat{\bGamma} - \bGamma^\star\|_F^2
    \le &
    2\lambda_n \|\hat{\bGamma} - \bGamma^\star\|_F 
    +
    \frac{3\lambda_n}{2}\|\Pi(\hat{\bGamma} - \bGamma^\star)\|_*
    +
    2\lambda_n\|\Pi^\perp(\bGamma^\star)\|_* \\
    \le &3\sqrt{2r}\lambda_n\|\hat{\bGamma} - \bGamma^\star\|_F + 2\lambda_n\|\Pi^\perp(\bGamma^\star)\|_* .  
\end{align}
Combining the choice of $\tau$ in Case 1, \eqref{eqn:r-inequality} and \eqref{eqn:Pi-perp-norm}, with Young's inequality, we have
\begin{align}
    \|\hat{\bGamma} - \bGamma^\star\|_F^2 \le CR_q\lambda_n^{2-q}. 
\end{align}

\textbf{Step 4. Justify the RSC condition.} It remains now to justify the RSC condition. We prove the following result holds with high probability:
\begin{align}
    \frac{1}{\sqrt{n}}\|\fS(\bDelta)\|_2 
    \ge 
    \frac{1}{8}\|\bDelta\|_F 
    -
    \frac{48 d \|\bDelta\|_\infty}{\sqrt{n}}
\end{align}
for all $\bDelta\in \bbR^{d_1 \times d_2}$ in the set  
\begin{align}
\cC = \left\{
\bDelta\in \bbR^{d_1\times d_2}  \mid 
    \|\bDelta\|_F^2 
    \ge
    c_0 (\sqrt{d_1d_2}\|\bDelta\|_\infty)
    \|\bDelta\|_*
    \sqrt{\frac{d\log d}{n}}\right\}. 
\end{align}
Define the following bad event: 
\begin{align}
    \cB 
    =
    \left\{
    \exists \bDelta\in\cC \mid
    \left| \frac{1}{\sqrt{n}}\|\fS(\bDelta)\|_2 - \|\bDelta\|_F \right|
    >
    \frac{7}{8}\|\bDelta\|_F
    +
    \frac{48 L d \|\bDelta\|_\infty}{\sqrt{n}} 
    \right\}. 
\end{align}
Now we peel the set $\cC$ with different radius:
\begin{align}
    \cC(D) = \{\bDelta \in \cC\mid \|\bDelta\|_\infty = d^{-1}, \|\bDelta\|_F\le D, 
    \|\bDelta\|_* \le D^2/(c_0\sqrt{d\log d/n})\}
\end{align}
and the peeled event 
\begin{align}
    \cB(D)
    =
    \lt\{\exists \bDelta \in \cC(D)
    \mid 
    \lt|\frac{1}{\sqrt{n}}\|\fS(\bGamma)\|_2 - \|\bDelta\|_F\rt| \ge \frac{3}{4}D + \frac{48L}{\sqrt{n}}
    \rt \}. 
\end{align}
By a discretization argument, Lemma 4 and Lemma 5 in \citet{negahban2012restricted} proved the following bound: with probability at least $1 - 4\exp(-cnD^2)$, it holds that 
\begin{align}
    \sup_{\bDelta \in \cB(D)}
    \lt|\frac{1}{\sqrt{n}}\|\fS(\bDelta)\|_2 - \|\bDelta\|_F\rt| \le \frac{3}{4}D + \frac{48L}{\sqrt{n}}. 
\end{align}
Now observing that for any $\bDelta \in \cC$ with $\|\bDelta\|_\infty = d^{-1}$,
we have that 
\begin{align}
    \|\bDelta\|_F^2 
    \ge
    c_0 (\sqrt{d_1d_2}\|\bDelta\|_\infty)\|\bDelta\|_* \sqrt{\frac{d\log d}{n}}
    \ge
    c_0\|\bDelta\|_* \sqrt{\frac{d\log d}{n}}
    \ge 
    c_0\|\bDelta\|_F \sqrt{\frac{d\log d}{n}}. 
\end{align}
This then leads to 
\begin{align}
    \|\bDelta\|_F \ge c_0 \sqrt{\frac{d\log d}{n}}. 
\end{align}
Therefore, we only need to focus on $\cC'$ where 
\begin{align}
    \cC' = 
    \{\bDelta \in \cC\mid \|\bDelta\| = d^{-1}, \|\bDelta\|_F \ge c_0\sqrt{d\log d/n}\}. 
\end{align}
By setting $\alpha = 7/6$ and $v = c_0\sqrt{d\log d/n}$, we can peel $\cC'$ into the following layers:  
\begin{align}
    \cC_l' = 
    \{\bDelta \in \cC \mid \|\bDelta\|_\infty = d^{-1}, 
    \alpha^{l-1} v
    \le \|\bDelta\|_F
    \le 
    \alpha^l v, 
    \|\bDelta\|_* \le 
    (\alpha^l v)^2/(c_0\sqrt{d\log d/n})\}.
\end{align}
If the bad event $\cB$ holds for some matrix $\bDelta$, then $\bDelta$ must belong to some set $\cC_l'$, which satisfies 
\begin{align}
    \lt|\frac{1}{\sqrt{n}}\|\fS(\bDelta)\|_2 - \|\bDelta\|_F\rt|
    >
    \frac{7}{8} \alpha^{l-1}v 
    +
    \frac{48L}{\sqrt{n}}
    =
    \frac{3}{4}\alpha^{l}v
    +
    \frac{48L}{\sqrt{n}}.
\end{align}
Thus, $\cB(\alpha^l v)$ must hold. Applying a union bound, we must have
\begin{align}
    \Prob{\cB} 
    \le &
    \sum_{l=1}^\infty
    \Prob{\cB(\alpha^l v)}
    \le 
    c_1 \sum_{l=1}^\infty
    \exp(-c_2 n \alpha^{2l}v^2)\\
    \le &
    c_1 \sum_{l=1}^\infty
    \exp(-2c_2 n \log(\alpha) l d\log d) \\
    = & 
    \frac{c_1 \exp(-c_2' d\log d)}{1 - \exp(-c_2'd\log d)} \\
    \lesssim & 
    c_1' \exp(-c_2'd\log d). 
\end{align}\end{proof}

\subsection{Proof of Theorem \ref{thm:regret}}
\begin{proof}[Proof of Theorem \ref{thm:regret}]
    
In the first stage, the regret is bounded by 
\begin{align}
    \text{Regret}_e
    =
    \sum_{t=1}^{T_e} 
    e^{\top}_{z^\star,t} \bGamma^\star 
    e_{x,t}
    -
    e^{\top}_{\hat{z},t}  \bGamma^\star e_{x,t}
    \le 
    2 T_e \cdot  \|\bGamma^\star\|_\infty.  
\end{align}

In the second stage, the regret is
\begin{align}
    \text{Regret}_c 
    = & \sum_{t=T_e + 1}^{T}
    e^{\star\top}_{z,t}
     \bGamma^\star 
    e^{\star}_{x,t}
    -
    \hat{e}^\top_{z,t}
      \bGamma^\star e^{\star}_{x,t}\\
    = & \sum_{t=T_e + 1}^{T}
    (e^{\star}_{z,t} - \hat{e}_{z,t})^\top  (\bGamma^\star - \hat{\bGamma}) e^\star_{x,t}  
    +
    (e^{\star}_{z,t} - \hat{e}_{z,t})^\top  \hat{\bGamma} e^\star_{x,t}
    \\
    \le & \sum_{t=T_e + 1}^{T}
    (e^{\star}_{z,t} - \hat{e}_{z,t})^\top  (\bGamma^\star - \hat{\bGamma}) e^\star_{x,t} \\ 
    \le & 2 T_c \cdot  \|\bGamma^\star - \hat{\bGamma}\|_\infty \\
    \le & 2 T_c \cdot \|\bGamma^\star - \hat{\bGamma}\|_F \\
    \le & 2 T_c \lt(\frac{d_1d_2\log(d_1 + d_2)r d}{T_e}\rt)^{1/2}.  
\end{align}
Therefore, the total regret across the two stages is 
\begin{align}
    \text{Regret} 
    = &
    \text{Regret}_e + \text{Regret}_c \\
    \le &
    2 T_e \cdot \|\bGamma^\star\|_\infty
    +
    2 (T - T_e) \cdot \|\bGamma^\star\|_\infty\cdot \lt(\frac{d_1d_2\log(d_1 + d_2)r d}{T_e}\rt)^{1/2}. 
\end{align}
Now setting $T_e = T^a$ for some $a\in(0,1)$. The above upper bound is minimized at 
\begin{align}
    a \asymp \frac{1}{3}\log_T \{C T^2 d_1d_2\log(d_1 + d_2)r d\}, 
\end{align}
which leads to a final regret bound
\begin{align}
    \text{Regret} 
    \le C T^{2/3} d\log d^{1/3} r^{1/3}. 
\end{align}
\end{proof}

\section{Additional simulation results}\label{sec:more-simulation}

\subsection{Comparison of several kernel-based estimation strategies} 
\begin{figure}[ht!]
    \centering
    \includegraphics[width=0.5\linewidth]{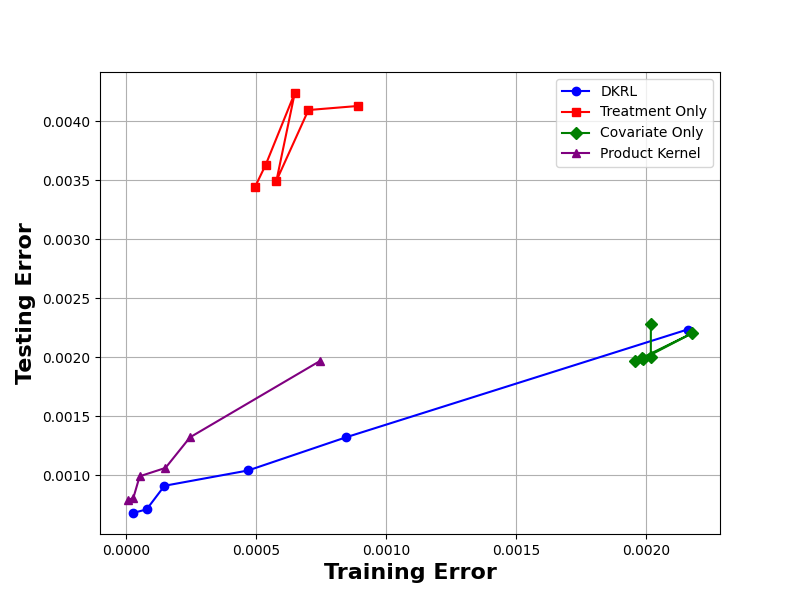}
    \caption{Training and testing error for different methods}
    \label{fig:train-test}
\end{figure}

Based on the setup of the Upworthy experiment, we also conduct a Monte Carlo simulation to evaluate the training and testing error of the proposed method. For the first experiment, we evaluate the training and testing error of the proposed method and compare it with several baseline methods. To do this, we split the synthetic dataset into a training set and a testing set, fit the training data with different methods, and evaluate the fitted results on the testing dataset. The four baseline methods we evaluate are: (i) double kernel representation learning (\texttt{DKRL}); (ii) RKHS regression with only treatments (\texttt{Treatment Only}); (iii) RKHS regression with only covariates (\texttt{Covariate Only}); (iv) RKHS regression with both treatments and covariates using a product kernel $\cK_g \odot \cK_h$ (\texttt{Product Kernel}). In our simulation, we vary across multiple levels of regularization parameters and compute the mean squared training and testing error.
The results are summarized in Figure \ref{fig:train-test}. 

\subsection{Interpretability of DKRL results}
To further elaborate on the interpretability of the learned feature, we show several example headlines with high or low semantic scores in Table \ref{tab:upworthy-interpretability}, which are good demonstrations on how the features capture the semantic meanings. 

On the X row representing "sharp contrasts with narrative surprises", high-scored examples carry stronger “contrast + surprise” hooks. Each headline pairs a familiar setup with a jarring twist—Hank Green sparring with a “corporate suit” over a supposedly dull yet vital topic, a Black cop repeatedly stop-and-frisked, and cute animals foreshadowing a darker reveal—creating instant narrative surprise. The low-scored examples, on the contrary, lean more on curiosity or emotional uplift (e.g., “A Lot Of You Will Want To Know About It Anyway”, “hese Pics Of A Baby Bird’s Rescue Are So Moving”) and less on jarring juxtapositions. They intrigue, but they don’t rely as heavily on opposites twists.

On the Y row representing "conflict-driven tone", high-scored examples signal confrontation, outrage, or direct criticism, with words like “Deport THIS GUY”, “attacks”, “smackdown”, “bad-ass”, “tears into the awful folks”. Besides, the examples pit one party against another (Hank Green vs. Corporate Suit, pink gang vs. predators, Fox Anchor vs. colleagues). The low-scored examples lean toward curiosity, wonder, and mild social critique. It uses softer hooks (“made me the happiest”, “reason to think twice”) and lacks the adversarial or combative language that characterizes the high-scored ones. 

\begin{table}[ht!]
\centering
\caption{Interpretability of the learned representations from DKRL in the Upworthy experiment. Feature X (X-axis in Figure \ref{fig:upworthy}(a)) represents whether there are "sharp contrasts with narrative surprises", and feature Y (Y-axis in Figure \ref{fig:upworthy}(a)) represents whether the sentence is in a "conflict-driven tone".}
\label{tab:upworthy-interpretability}
\begin{tabular}{P{1cm}L{5.8cm}L{5.8cm}}
\toprule
\textbf{Feature} & \multicolumn{1}{c}{\textbf{Example headlines with high scores}}  & \multicolumn{1}{c}{\textbf{Example headlines with low scores}}  \\
\midrule
X &
- “Hank Green Goes Up Against A Corporate Suit To Debate An Unsexy Issue That Is Really A Big Deal”; \par\medskip

- “He’s Black. He’s a Cop. He’s Also Been Stopped and Frisked 30 Times.”;  \par\medskip

- “Floppy Ears, Soulful Eyes, Adorable Babies. And They Have Something Else People Can’t Stop Taking.” &

- “Anyone Able To Read This Won't Benefit From It, But A Lot Of You Will Want To Know About It Anyway”; \par\medskip

- “She Won A Golden Globe And Dedicated It To A Teen She Never Met. Here's Why It's A Big Deal.”;\par\medskip

- “These Pics Of A Baby Bird's Rescue Are So Moving You Might Get Emotionally Involved Here” \\
\midrule 
Y &
- “Hey, I Have An Idea: Let's Deport THIS GUY, Instead”; \par\medskip

- “This Pink Gang Attacks Sexual Predators And Rapists That Harm Them. Tell Me, What Else Can They Do?”; \par\medskip

- “PLOT TWIST: Fox News Anchor Lays Epic Smackdown On Fox News Anchors For Obvious And Blatant Misogyny” &

- “A Tiny Probe Millions Of Miles Away Just Made Me The Happiest I've Been In Weeks”; \par\medskip

- “Here’s A List Of Uncool Problems For Ladies That We Could Fix To Make Life Easier For Everyone”; \par\medskip

- “A Reason To Think Twice Before Buying Your Kid All Those LEGOs” \\
\bottomrule 
\end{tabular}
\end{table}

\subsection{Sensitivity to hyperparameter tuning}

\begin{figure}[ht]
     \centering
     \includegraphics[width=0.6\linewidth]{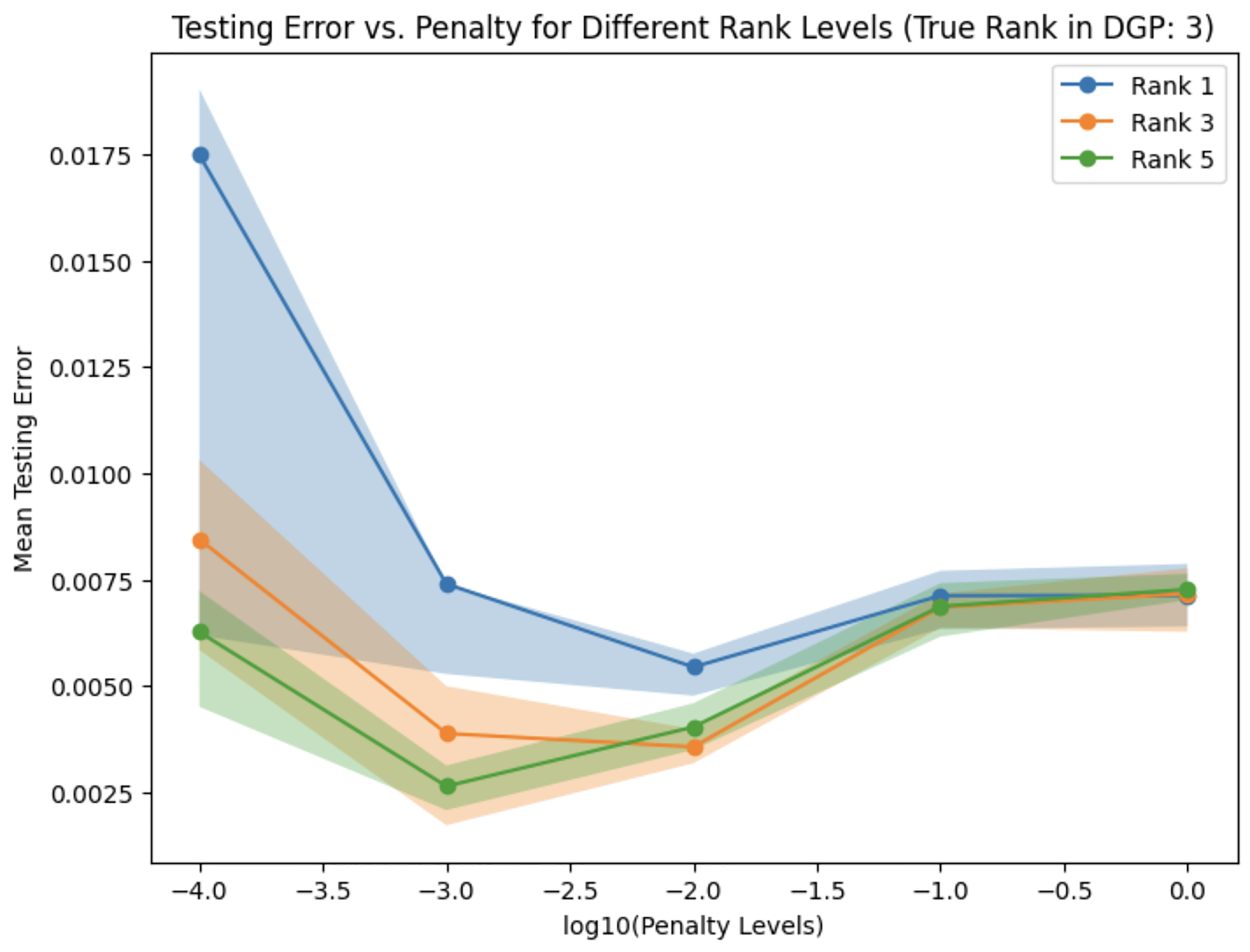}
     \caption{Hyperparameter sensitivity against penalty level $\lambda$ and rank $r$. }
     \label{fig:sensitivity}
 \end{figure}

We evaluated the sensitivity of our Algorithm \ref{alg:UV} to two tuning parameters, penalty level $\lambda$ and rank $r$, in Algorithm \ref{alg:UV}. The results are reported in Figure \ref{fig:sensitivity}. In terms of selection of penalty level, an appropriate choice of penalty is important to balance the level of regularization and loss minimization. In practice, we recommend using cross validation to tune this parameter. In terms of rank selection, we can see that selecting a rank that is smaller than the truth could lead to bias, as this loses important features. In contrast, there is minimal bias from selecting a slightly larger rank. In practice, we recommend starting with a larger rank and trimming slightly.

\subsection{The MIND dataset experiment}
In this subsection, we report the results for the synthetic simulation we conducted for the MIND dataset in Table \ref{tab:RMSE-MIND}. As we have commented in the main paper, while most methods benefit from the low-rank structure, DKRL exploits the low-rank structure more effectively. And even with a high-rank interaction matrix, DKRL utilizes the similarity structure in the embeddings to enhance accuracy. 

\begin{table}[ht!]
\centering
\caption{Testing MSE across methods (rows) and decay levels (columns)}
\label{tab:RMSE-MIND}
\begin{tabular}{l cccc}
\toprule
Method & $q=0$ & $q=2$ & $q=4$ & $q=6$ \\
\midrule
DKRL      & \textbf{0.007 (0.003)} & \textbf{0.003 (0.001)} & \textbf{0.003 (0.001)} & \textbf{0.003 (0.001)} \\
Lasso\_Z  & 0.034 (0.003) & 0.009 (0.002) & 0.008 (0.002) & 0.008 (0.002) \\
XGB\_Z    & 0.036 (0.003) & 0.007 (0.002) & 0.007 (0.002) & 0.007 (0.002) \\
FNN\_Z    & 0.034 (0.003) & 0.011 (0.002) & 0.011 (0.002) & 0.011 (0.002) \\
Kernel\_Z & 0.031 (0.003) & 0.006 (0.002) & 0.006 (0.001) & 0.006 (0.002) \\
Lasso\_X  & 0.055 (0.005) & 0.013 (0.004) & 0.011 (0.003) & 0.011 (0.003) \\
XGB\_X    & 0.056 (0.006) & 0.014 (0.004) & 0.011 (0.003) & 0.011 (0.003) \\
FNN\_X    & 0.055 (0.005) & 0.014 (0.004) & 0.011 (0.003) & 0.011 (0.003) \\
Kernel\_X & 0.055 (0.005) & 0.013 (0.004) & 0.011 (0.003) & 0.011 (0.003) \\
Lasso\_ZX & 0.032 (0.003) & 0.009 (0.001) & 0.008 (0.002) & 0.008 (0.002) \\
XGB\_ZX   & 0.022 (0.001) & 0.004 (0.001) & 0.004 (0.001) & 0.004 (0.001) \\
FNN\_ZX   & 0.016 (0.002) & 0.010 (0.002) & 0.009 (0.002) & 0.010 (0.002) \\
Kernel\_ZX& 0.035 (0.002) & 0.008 (0.002) & 0.007 (0.001) & 0.007 (0.002) \\
\bottomrule
\end{tabular}
\end{table}

\subsection{The ASOS experiment}
In the second simulation based on the ASOS experiment, we generated a synthetic study following the similar DGP to the one used in the main paper, with newly generated embeddings from the shopping item information from the ASOS dataset. We then compared the KDRL method with two other methods: deep-learning based (SIN) method and product-kernel based (ProdKernel) method in terms of their training/testing error and computation time. The results are reported in Table \ref{tab:asos-tab}, from which we consolidate the finding that DKRL performs fairly satisfactorily in terms of both training/testing errors and computation time.
\begin{table}[ht]
\centering
\caption{ASOS Experiment by Rank and Method (Mean (Std), rounded to four decimals)}
\label{tab:asos-tab}
\begin{tabular}{llccc}
\toprule
\textbf{Rank} & \textbf{Method}         & \textbf{Train Error}             & \textbf{Test Error}              & \textbf{Time}                  \\
\midrule
\multirow{3}{*}{2} 
  & \texttt{SIN}            & 0.0094 (0.0004)         & 0.0095 (0.0011)         & 7.9361 (0.4858)       \\
  & \texttt{DKRL}           & \textbf{0.0022 (0.0002)}         & \textbf{0.0035 (0.0009)}         & 0.1739 (0.0305)       \\
  & \texttt{ProdKernel} & 0.0036 (0.0001)         & 0.0062 (0.0010)         & 0.0013 (0.0002)       \\
\midrule
\multirow{3}{*}{3} 
  & \texttt{SIN}           & 0.0085 (0.0004)         & 0.0085 (0.0011)         & 8.0929 (0.3153)       \\
  & \texttt{DKRL}           & \textbf{0.0025 (0.0002)}         & \textbf{0.0047 (0.0010)}         & 0.4702 (0.1094)       \\
  & \texttt{ProdKernel} & 0.0044 (0.0002)         & 0.0075 (0.0012)         & 0.0014 (0.0003)       \\
\midrule
\multirow{3}{*}{5} 
  & \texttt{SIN}            & 0.0119 (0.0014)         & 0.0120 (0.0014)         & 7.9983 (0.4951)       \\
  & \texttt{DKRL}           & \textbf{0.0027 (0.0000)}         & \textbf{0.0076 (0.0012)}         & 1.6868 (0.3907)       \\
  & \texttt{ProdKernel} & 0.0057 (0.0000)         & 0.0100 (0.0014)         & 0.0014 (0.0004)       \\
\midrule
\multirow{3}{*}{10} 
  & \texttt{SIN}            & 0.0175 (0.0007)         & 0.0175 (0.0018)         & 8.1151 (0.4709)       \\
  & \texttt{DKRL}           & \textbf{0.0026 (0.0001)}         & \textbf{0.0128 (0.0017)}         & 7.9518 (0.8747)       \\
  & \texttt{ProdKernel} & 0.0083 (0.0003)         & 0.0147 (0.0017)         & 0.0017 (0.0004)       \\
\bottomrule
\end{tabular}
\end{table}


\bigskip


\newpage
\section*{NeurIPS Paper Checklist}

\begin{enumerate}

\item {\bf Claims}
    \item[] Question: Do the main claims made in the abstract and introduction accurately reflect the paper's contributions and scope?
    \item[] Answer: \answerYes{} 
    \item[] Justification: we outlined our contributions in the abstract and Section \ref{sec:intro}. Section \ref{sec:problem}-\ref{sec:simulation} fully elaborates on all the points we made. 
    
    \item[] Guidelines:
    \begin{itemize}
        \item The answer NA means that the abstract and introduction do not include the claims made in the paper.
        \item The abstract and/or introduction should clearly state the claims made, including the contributions made in the paper and important assumptions and limitations. A No or NA answer to this question will not be perceived well by the reviewers. 
        \item The claims made should match theoretical and experimental results, and reflect how much the results can be expected to generalize to other settings. 
        \item It is fine to include aspirational goals as motivation as long as it is clear that these goals are not attained by the paper. 
    \end{itemize}

\item {\bf Limitations}
    \item[] Question: Does the paper discuss the limitations of the work performed by the authors?
    \item[] Answer: \answerYes{} 
    \item[] Justification: we discussed the limitation of the work in Conclusion (Section \ref{sec:conclusion}).  
    \item[] Guidelines:
    \begin{itemize}
        \item The answer NA means that the paper has no limitation while the answer No means that the paper has limitations, but those are not discussed in the paper. 
        \item The authors are encouraged to create a separate "Limitations" section in their paper.
        \item The paper should point out any strong assumptions and how robust the results are to violations of these assumptions (e.g., independence assumptions, noiseless settings, model well-specification, asymptotic approximations only holding locally). The authors should reflect on how these assumptions might be violated in practice and what the implications would be.
        \item The authors should reflect on the scope of the claims made, e.g., if the approach was only tested on a few datasets or with a few runs. In general, empirical results often depend on implicit assumptions, which should be articulated.
        \item The authors should reflect on the factors that influence the performance of the approach. For example, a facial recognition algorithm may perform poorly when image resolution is low or images are taken in low lighting. Or a speech-to-text system might not be used reliably to provide closed captions for online lectures because it fails to handle technical jargon.
        \item The authors should discuss the computational efficiency of the proposed algorithms and how they scale with dataset size.
        \item If applicable, the authors should discuss possible limitations of their approach to address problems of privacy and fairness.
        \item While the authors might fear that complete honesty about limitations might be used by reviewers as grounds for rejection, a worse outcome might be that reviewers discover limitations that aren't acknowledged in the paper. The authors should use their best judgment and recognize that individual actions in favor of transparency play an important role in developing norms that preserve the integrity of the community. Reviewers will be specifically instructed to not penalize honesty concerning limitations.
    \end{itemize}

\item {\bf Theory assumptions and proofs}
    \item[] Question: For each theoretical result, does the paper provide the full set of assumptions and a complete (and correct) proof?
    \item[] Answer: \answerYes{} 
    \item[] Justification: we presented the full sets of assumptions in the statement of the Theorems (Theorem \ref{thm:representor}, \ref{thm:hGamma-bound}, \ref{thm:regret}), and provide proofs to all the claims we made in Appendix \ref{sec:proof}.
    \item[] Guidelines:
    \begin{itemize}
        \item The answer NA means that the paper does not include theoretical results. 
        \item All the theorems, formulas, and proofs in the paper should be numbered and cross-referenced.
        \item All assumptions should be clearly stated or referenced in the statement of any theorems.
        \item The proofs can either appear in the main paper or the supplemental material, but if they appear in the supplemental material, the authors are encouraged to provide a short proof sketch to provide intuition. 
        \item Inversely, any informal proof provided in the core of the paper should be complemented by formal proofs provided in appendix or supplemental material.
        \item Theorems and Lemmas that the proof relies upon should be properly referenced. 
    \end{itemize}

    \item {\bf Experimental result reproducibility}
    \item[] Question: Does the paper fully disclose all the information needed to reproduce the main experimental results of the paper to the extent that it affects the main claims and/or conclusions of the paper (regardless of whether the code and data are provided or not)?
    \item[] Answer: \answerYes{} 
    \item[] Justification: we include the anonymized version of our code as supplementary materials and we are happy to release the code and data once accepted. Our main experimental results mainly involve an implementation of our algorithms, which is detailed in \ref{alg:UV} and \ref{alg:etc}, and running them on synthetic datasets which are publicly available as detailed in Section \ref{sec:simulation} of the paper. 
    \item[] Guidelines:
    \begin{itemize}
        \item The answer NA means that the paper does not include experiments.
        \item If the paper includes experiments, a No answer to this question will not be perceived well by the reviewers: Making the paper reproducible is important, regardless of whether the code and data are provided or not.
        \item If the contribution is a dataset and/or model, the authors should describe the steps taken to make their results reproducible or verifiable. 
        \item Depending on the contribution, reproducibility can be accomplished in various ways. For example, if the contribution is a novel architecture, describing the architecture fully might suffice, or if the contribution is a specific model and empirical evaluation, it may be necessary to either make it possible for others to replicate the model with the same dataset, or provide access to the model. In general. releasing code and data is often one good way to accomplish this, but reproducibility can also be provided via detailed instructions for how to replicate the results, access to a hosted model (e.g., in the case of a large language model), releasing of a model checkpoint, or other means that are appropriate to the research performed.
        \item While NeurIPS does not require releasing code, the conference does require all submissions to provide some reasonable avenue for reproducibility, which may depend on the nature of the contribution. For example
        \begin{enumerate}
            \item If the contribution is primarily a new algorithm, the paper should make it clear how to reproduce that algorithm.
            \item If the contribution is primarily a new model architecture, the paper should describe the architecture clearly and fully.
            \item If the contribution is a new model (e.g., a large language model), then there should either be a way to access this model for reproducing the results or a way to reproduce the model (e.g., with an open-source dataset or instructions for how to construct the dataset).
            \item We recognize that reproducibility may be tricky in some cases, in which case authors are welcome to describe the particular way they provide for reproducibility. In the case of closed-source models, it may be that access to the model is limited in some way (e.g., to registered users), but it should be possible for other researchers to have some path to reproducing or verifying the results.
        \end{enumerate}
    \end{itemize}

\item {\bf Open access to data and code}
    \item[] Question: Does the paper provide open access to the data and code, with sufficient instructions to faithfully reproduce the main experimental results, as described in supplemental material?
    \item[] Answer: \answerYes{} 
    \item[] Justification: We submit an anonymized version of our code as supplementary materials.
    \item[] Guidelines:
    \begin{itemize}
        \item The answer NA means that paper does not include experiments requiring code.
        \item Please see the NeurIPS code and data submission guidelines (\url{https://nips.cc/public/guides/CodeSubmissionPolicy}) for more details.
        \item While we encourage the release of code and data, we understand that this might not be possible, so “No” is an acceptable answer. Papers cannot be rejected simply for not including code, unless this is central to the contribution (e.g., for a new open-source benchmark).
        \item The instructions should contain the exact command and environment needed to run to reproduce the results. See the NeurIPS code and data submission guidelines (\url{https://nips.cc/public/guides/CodeSubmissionPolicy}) for more details.
        \item The authors should provide instructions on data access and preparation, including how to access the raw data, preprocessed data, intermediate data, and generated data, etc.
        \item The authors should provide scripts to reproduce all experimental results for the new proposed method and baselines. If only a subset of experiments are reproducible, they should state which ones are omitted from the script and why.
        \item At submission time, to preserve anonymity, the authors should release anonymized versions (if applicable).
        \item Providing as much information as possible in supplemental material (appended to the paper) is recommended, but including URLs to data and code is permitted.
    \end{itemize}

\item {\bf Experimental setting/details}
    \item[] Question: Does the paper specify all the training and test details (e.g., data splits, hyperparameters, how they were chosen, type of optimizer, etc.) necessary to understand the results?
    \item[] Answer: \answerYes{} 
    \item[] Justification: These are discussed in Section \ref{sec:simulation} of the main paper. 
    \item[] Guidelines:
    \begin{itemize}
        \item The answer NA means that the paper does not include experiments.
        \item The experimental setting should be presented in the core of the paper to a level of detail that is necessary to appreciate the results and make sense of them.
        \item The full details can be provided either with the code, in appendix, or as supplemental material.
    \end{itemize}

\item {\bf Experiment statistical significance}
    \item[] Question: Does the paper report error bars suitably and correctly defined or other appropriate information about the statistical significance of the experiments?
    \item[] Answer: \answerYes{} 
    \item[] Justification: we include standard deviations or inter-quantile ranges for the experiment results. 
    \item[] Guidelines:
    \begin{itemize}
        \item The answer NA means that the paper does not include experiments.
        \item The authors should answer "Yes" if the results are accompanied by error bars, confidence intervals, or statistical significance tests, at least for the experiments that support the main claims of the paper.
        \item The factors of variability that the error bars are capturing should be clearly stated (for example, train/test split, initialization, random drawing of some parameter, or overall run with given experimental conditions).
        \item The method for calculating the error bars should be explained (closed form formula, call to a library function, bootstrap, etc.)
        \item The assumptions made should be given (e.g., Normally distributed errors).
        \item It should be clear whether the error bar is the standard deviation or the standard error of the mean.
        \item It is OK to report 1-sigma error bars, but one should state it. The authors should preferably report a 2-sigma error bar than state that they have a 96\% CI, if the hypothesis of Normality of errors is not verified.
        \item For asymmetric distributions, the authors should be careful not to show in tables or figures symmetric error bars that would yield results that are out of range (e.g. negative error rates).
        \item If error bars are reported in tables or plots, The authors should explain in the text how they were calculated and reference the corresponding figures or tables in the text.
    \end{itemize}

\item {\bf Experiments compute resources}
    \item[] Question: For each experiment, does the paper provide sufficient information on the computer resources (type of compute workers, memory, time of execution) needed to reproduce the experiments?
    \item[] Answer: \answerYes{} 
    \item[] Justification: we used personal computers (MBP with a M2 Max chip) to run the experiments on CPU and did not use additional computing resources. This information is included in Section \ref{sec:simulation}. 
    \item[] Guidelines:
    \begin{itemize}
        \item The answer NA means that the paper does not include experiments.
        \item The paper should indicate the type of compute workers CPU or GPU, internal cluster, or cloud provider, including relevant memory and storage.
        \item The paper should provide the amount of compute required for each of the individual experimental runs as well as estimate the total compute. 
        \item The paper should disclose whether the full research project required more compute than the experiments reported in the paper (e.g., preliminary or failed experiments that didn't make it into the paper). 
    \end{itemize}
    
\item {\bf Code of ethics}
    \item[] Question: Does the research conducted in the paper conform, in every respect, with the NeurIPS Code of Ethics \url{https://neurips.cc/public/EthicsGuidelines}?
    \item[] Answer: \answerYes{} 
    \item[] Justification: we follow the NeurIPS Code of Ethics. 
    \item[] Guidelines:
    \begin{itemize}
        \item The answer NA means that the authors have not reviewed the NeurIPS Code of Ethics.
        \item If the authors answer No, they should explain the special circumstances that require a deviation from the Code of Ethics.
        \item The authors should make sure to preserve anonymity (e.g., if there is a special consideration due to laws or regulations in their jurisdiction).
    \end{itemize}

\item {\bf Broader impacts}
    \item[] Question: Does the paper discuss both potential positive societal impacts and negative societal impacts of the work performed?
    \item[] Answer: \answerYes{} 
    \item[] Justification: we outlined the social impacts of the paper in advancing the techniques for digital experimentation in our contribution. Yet there could negative impacts related to the use of Large Language Models in experimentation such as privacy and fairness issues. We also outlined this in the Conclusion section. 
    
    \item[] Guidelines:
    \begin{itemize}
        \item The answer NA means that there is no societal impact of the work performed.
        \item If the authors answer NA or No, they should explain why their work has no societal impact or why the paper does not address societal impact.
        \item Examples of negative societal impacts include potential malicious or unintended uses (e.g., disinformation, generating fake profiles, surveillance), fairness considerations (e.g., deployment of technologies that could make decisions that unfairly impact specific groups), privacy considerations, and security considerations.
        \item The conference expects that many papers will be foundational research and not tied to particular applications, let alone deployments. However, if there is a direct path to any negative applications, the authors should point it out. For example, it is legitimate to point out that an improvement in the quality of generative models could be used to generate deepfakes for disinformation. On the other hand, it is not needed to point out that a generic algorithm for optimizing neural networks could enable people to train models that generate Deepfakes faster.
        \item The authors should consider possible harms that could arise when the technology is being used as intended and functioning correctly, harms that could arise when the technology is being used as intended but gives incorrect results, and harms following from (intentional or unintentional) misuse of the technology.
        \item If there are negative societal impacts, the authors could also discuss possible mitigation strategies (e.g., gated release of models, providing defenses in addition to attacks, mechanisms for monitoring misuse, mechanisms to monitor how a system learns from feedback over time, improving the efficiency and accessibility of ML).
    \end{itemize}
    
\item {\bf Safeguards}
    \item[] Question: Does the paper describe safeguards that have been put in place for responsible release of data or models that have a high risk for misuse (e.g., pretrained language models, image generators, or scraped datasets)?
    \item[] Answer: \answerNA{} 
    \item[] Justification: the paper poses no such risks. 
    \item[] Guidelines:
    \begin{itemize}
        \item The answer NA means that the paper poses no such risks.
        \item Released models that have a high risk for misuse or dual-use should be released with necessary safeguards to allow for controlled use of the model, for example by requiring that users adhere to usage guidelines or restrictions to access the model or implementing safety filters. 
        \item Datasets that have been scraped from the Internet could pose safety risks. The authors should describe how they avoided releasing unsafe images.
        \item We recognize that providing effective safeguards is challenging, and many papers do not require this, but we encourage authors to take this into account and make a best faith effort.
    \end{itemize}

\item {\bf Licenses for existing assets}
    \item[] Question: Are the creators or original owners of assets (e.g., code, data, models), used in the paper, properly credited and are the license and terms of use explicitly mentioned and properly respected?
    \item[] Answer: \answerNA{} 
    \item[] Justification: the paper does not use existing assets. 
    \item[] Guidelines:
    \begin{itemize}
        \item The answer NA means that the paper does not use existing assets.
        \item The authors should cite the original paper that produced the code package or dataset.
        \item The authors should state which version of the asset is used and, if possible, include a URL.
        \item The name of the license (e.g., CC-BY 4.0) should be included for each asset.
        \item For scraped data from a particular source (e.g., website), the copyright and terms of service of that source should be provided.
        \item If assets are released, the license, copyright information, and terms of use in the package should be provided. For popular datasets, \url{paperswithcode.com/datasets} has curated licenses for some datasets. Their licensing guide can help determine the license of a dataset.
        \item For existing datasets that are re-packaged, both the original license and the license of the derived asset (if it has changed) should be provided.
        \item If this information is not available online, the authors are encouraged to reach out to the asset's creators.
    \end{itemize}

\item {\bf New assets}
    \item[] Question: Are new assets introduced in the paper well documented and is the documentation provided alongside the assets?
    \item[] Answer: \answerNA{} 
    \item[] Justification: the paper does not release new assets. 
    \item[] Guidelines:
    \begin{itemize}
        \item The answer NA means that the paper does not release new assets.
        \item Researchers should communicate the details of the dataset/code/model as part of their submissions via structured templates. This includes details about training, license, limitations, etc. 
        \item The paper should discuss whether and how consent was obtained from people whose asset is used.
        \item At submission time, remember to anonymize your assets (if applicable). You can either create an anonymized URL or include an anonymized zip file.
    \end{itemize}

\item {\bf Crowdsourcing and research with human subjects}
    \item[] Question: For crowdsourcing experiments and research with human subjects, does the paper include the full text of instructions given to participants and screenshots, if applicable, as well as details about compensation (if any)? 
    \item[] Answer: \answerNA{} 
    \item[] Justification: the paper does not involve crowdsourcing nor research with human subjects. 
    \item[] Guidelines:
    \begin{itemize}
        \item The answer NA means that the paper does not involve crowdsourcing nor research with human subjects.
        \item Including this information in the supplemental material is fine, but if the main contribution of the paper involves human subjects, then as much detail as possible should be included in the main paper. 
        \item According to the NeurIPS Code of Ethics, workers involved in data collection, curation, or other labor should be paid at least the minimum wage in the country of the data collector. 
    \end{itemize}

\item {\bf Institutional review board (IRB) approvals or equivalent for research with human subjects}
    \item[] Question: Does the paper describe potential risks incurred by study participants, whether such risks were disclosed to the subjects, and whether Institutional Review Board (IRB) approvals (or an equivalent approval/review based on the requirements of your country or institution) were obtained?
    \item[] Answer: \answerNA{} 
    \item[] Justification: the paper does not involve crowdsourcing nor research with human subjects.  
    \item[] Guidelines:
    \begin{itemize}
        \item The answer NA means that the paper does not involve crowdsourcing nor research with human subjects.
        \item Depending on the country in which research is conducted, IRB approval (or equivalent) may be required for any human subjects research. If you obtained IRB approval, you should clearly state this in the paper. 
        \item We recognize that the procedures for this may vary significantly between institutions and locations, and we expect authors to adhere to the NeurIPS Code of Ethics and the guidelines for their institution. 
        \item For initial submissions, do not include any information that would break anonymity (if applicable), such as the institution conducting the review.
    \end{itemize}

\item {\bf Declaration of LLM usage}
    \item[] Question: Does the paper describe the usage of LLMs if it is an important, original, or non-standard component of the core methods in this research? Note that if the LLM is used only for writing, editing, or formatting purposes and does not impact the core methodology, scientific rigorousness, or originality of the research, declaration is not required.
    \item[] Answer: \answerYes{} 
    \item[] Justification: we discussed the use of LLMs as this is a core part of our methodology. 
    \item[] Guidelines:
    \begin{itemize}
        \item The answer NA means that the core method development in this research does not involve LLMs as any important, original, or non-standard components.
        \item Please refer to our LLM policy (\url{https://neurips.cc/Conferences/2025/LLM}) for what should or should not be described.
    \end{itemize}

\end{enumerate}

\end{document}